\documentclass[11pt,a4paper]{article}
\usepackage{jheppub}
\usepackage{amsmath,amssymb,amsfonts}
\usepackage{graphicx,graphics}
\usepackage{graphics}
\usepackage{url}
\usepackage{slashed}
\usepackage{color}
\usepackage{mathtools}
\usepackage[utf8]{inputenc}
\usepackage{subcaption}

\usepackage{siunitx}
\newcommand{\msbar}{\overline{\mathrm{MS}}}
\newcommand{\renorm}{\mu_{\text{ren}}}
\newcommand{\fact}{\mu_{\text{fact}}}
\newcommand{\frag}{\mu_{\text{frag}}}
\newcommand{\dd}{\mathrm{d}}
\newcommand{\Gammatot}{\Gamma_{\mathrm{tot}}}

\title{Dimuons from neutrino-nucleus collisions in the semi-inclusive DIS approach}

\affiliation{University of Jyväskylä, Department of Physics, P.O. Box 35, FI-40014 University of Jyväskylä, Finland}
\affiliation{Helsinki Institute of Physics, P.O. Box 64, FI-00014 University of Helsinki, Finland}

\emailAdd{ilkka.m.helenius@jyu.fi}
\emailAdd{hannu.paukkunen@jyu.fi}
\emailAdd{sami.a.yrjanheikki@jyu.fi}

\abstract{
We present a next-to-leading order perturbative QCD calculation of dimuon production in neutrino-nucleus collisions. This process is typically calculated by assuming it to be proportional to inclusive charm production, which requires an effective acceptance correction to take the experimental cuts on the decay-muon kinematics
into account. Here, we instead compute the dimuon production cross section directly as a convolution of semi-inclusive deep inelastic scattering to produce charmed hadrons, and a decay function fitted to $e^+e^-$ data to produce a muon from the charmed hadrons. The presented approach is in a good agreement with available experimental data and will serve as a starting point for higher-order QCD calculations without an external acceptance correction. The uncertainties arising from the decay function and scale dependence are sizeably smaller than those from the nuclear parton distribution functions. We also calculate the effective acceptances within our approach and compare them to those usually used in global fits of parton distribution functions, finding differences of the order of $10\,\%$, depending on the kinematics, perturbative order, and applied parton distributions.
}

\keywords{Dimuon production, muon-pair production, nuclear parton distribution functions, QCD, neutrino SIDIS}

\begin{document}

\author{Ilkka Helenius,}
\author{Hannu Paukkunen and}
\author{Sami Yrjänheikki}

\maketitle

%%%%%%%%%%%%%%%%%%%%%%%%%%%%%%%%%%%%%%%%%%%%%%%%%%%%%%%%%%%%%%%%%%%%%%%%%%%%%%%%%%%%%%%%%%%%%%%%%%%%%%%%%%%
\section{Introduction}
\label{sec:intro}

A staple of Quantum Chromodynamics (QCD) is collinear factorization \cite{Collins:1989gx}, which allows for the calculation of hadronic cross sections by separating the effects of long- and short-distance physics. The long-distance effects are described by universal, process-independent, parton distribution functions (PDFs) \cite{Kovarik:2019xvh,Ethier:2020way,Klasen:2023uqj} and fragmentation functions (FFs) \cite{Metz:2016swz}. While non-perturbative and thus not calculable from perturbative QCD, their scale dependence is still predicted by the perturbative Dokshitzer-Gribov-Lipatov-Altarelli-Parisi (DGLAP) evolution equations \cite{Gribov:1972ri,Gribov:1972rt,Dokshitzer:1977sg,Altarelli:1977zs}. The short-distance effects, on the other hand, are perturbatively calculable due to the asymptotic freedom of QCD.

One of the most poorly known components in the global analysis of free- and bound-proton (nuclear) PDFs is the strange-quark distribution \cite{Hou:2019efy,Bailey:2020ooq,NNPDF:2021njg,Eskola:2021nhw,Duwentaster:2022kpv,AbdulKhalek:2022fyi}. This in part limits e.g. the precision determination of Standard Model parameters --- like the mass of the $W^\pm$ boson \cite{ATLAS:2017rzl,Bagnaschi:2019mzi,Hussein:2019kqx,ATLAS:2023fsi} or the weak mixing angle \cite{CMS:2018ktx} --- from the data collected at the Large Hadron Collider (LHC). An important process to constrain the strange-quark content of nucleons is the semi-inclusive dimuon production in charged-current neutrino-nucleus ($\nu_\mu N$) deep-inelastic scattering (DIS) \cite{CCFR:1994ikl,NuTeV:2001dfo,NuTeV:2007uwm,CHORUS:2008vjb,NOMAD:2013hbk}. In comparison to the fully inclusive neutrino DIS, whose compatibility with the neutral-current DIS in the case of nuclear targets has been discussed quite thoroughly \cite{Schienbein:2009kk,Paukkunen:2010hb,Kovarik:2010uv,Paukkunen:2013grz,Muzakka:2022wey}, this process produces a charm quark in the partonic final state, which then hadronizes into a charmed hadron and eventually decays into a final state containing a secondary decay muon. Due to the elements of the Cabibbo–Kobayashi–Maskawa (CKM) matrix, $\left|V_{cs}\right|\approx 0.973$, $\left|V_{cd}\right|\approx 0.225$, and $\left|V_{cb}\right|\approx 0.042$ \cite{Workman:2022ynf}, the outgoing charm quark couples mostly to the strange-quark content of the target. While the neutrino data have been taken with heavy nuclear targets to obtain sufficient statistics, they are still typically used in global fits of free-proton PDFs \cite{Hou:2019efy,Bailey:2020ooq,NNPDF:2021njg}. This is taken into account by either ``correcting'' the data \cite{Hou:2019efy,Accardi:2021ysh} for the nuclear effects or treating them as theoretical uncertainties \cite{Ball:2018twp,Bailey:2020ooq,NNPDF:2021njg} tied to the corrections obtained from nuclear PDFs. Ultimately, this creates a correlation between proton and nuclear PDFs. In the future, experiments employing far-forward neutrinos originating from proton-proton collisions at the LHC could provide further constraints on the strange-quark distribution \cite{Cruz-Martinez:2023sdv}. Indeed, the first collider neutrinos have already been detected by the FASER \cite{FASER:2023zcr} and SND@LHC \cite{SNDLHC:2023pun} collaborations.

By including LHC proton-proton ($pp$) data into a global fit of PDFs, one can hope to reduce the dependence of the strange-quark PDFs on the treatment of nuclear corrections. While the inclusive $Z$ and $W^\pm$ boson production processes carry a significant contribution from channels involving a strange quark \cite{Kusina:2012vh}, they are still subleading in comparison to the contributions from other production channels involving up and down quarks. Moreover, at the high interaction scale $Q^2 \sim 10^4 \, {\rm GeV^2}$ relevant for these processes, a big part of the behaviour of the strange-quark PDFs is simply dictated by QCD with gluons splitting into $s\bar{s}$-pairs. As a result, it is not particularly easy to faithfully resolve the non-perturbative strange-quark density from such inclusive measurements \cite{Eskola:2022rlm} and e.g. any deficiencies in the theoretical description can manifest themselves as shifts in the strange-quark PDF just because it is much less constrained than e.g. the up- and down-quark densities \cite{Kusina:2020lyz}. The background from other partonic processes can be suppressed by considering the production of $W^\pm$ bosons associated with a charmed particle \cite{Baur:1993zd,Stirling:2012vh,ATLAS:2014jkm,CMS:2013wql,CMS:2018dxg,ATLAS:2023ibp}. However, the interaction scale is still high, and the charm identification increases the experimental uncertainties. Interestingly, the LHC $pp$ data have displayed some tensions with the neutrino dimuon data, the former preferring a larger ratio $R_s=(s+\bar{s})/(\bar{u}+\bar{d})$ at $x \approx 0.02$ \cite{ATLAS:2012sjl,ATLAS:2016nqi,Cooper-Sarkar:2018ufj,NNPDF:2017mvq,Hou:2019efy,Bailey:2020ooq}. However, higher-order QCD corrections on the dimuon cross sections tend to be negative \cite{Berger:2016inr,Gao:2017kkx} and appear to somewhat ease the observed tensions \cite{Faura:2020oom}. The strange-quark PDF can also be constrained by semi-inclusive kaon production in DIS \cite{HERMES:2013ztj,Borsa:2017vwy,Sato:2019yez} on protons, although in this case the uncertainties in the parton-to-kaon FFs limit the precision. As a result, semi-inclusive dimuon production still remains an integral part in constraining the stange-quark PDFs of the proton.

The semi-inclusive dimuon production cross section in neutrino-nucleus DIS is typically computed by assuming that it is proportional to the inclusive charm production cross section
\begin{equation}
\label{eq:charm_production_factorization}
	\dd\sigma(\nu_\mu N\to \mu^- \mu^+ X)\simeq\mathcal{A}\mathcal{B}_\mu \dd\sigma(\nu_\mu N\to \mu^- c X),
\end{equation}
where the acceptance correction $\mathcal{A}$ compensates for the experimental energy cut on the decay muon originating from the process and $\mathcal{B}_\mu$ is the semileptonic branching ratio averaged over charm mesons. At leading-order (LO) fixed-flavour number scheme (no heavy-quark PDFs), the dimuon cross section indeed factorizes as above, but at higher perturbative orders or in variable flavour-number scheme this is no longer the case. This poses a potential problem for increasing the accuracy of the dimuon production calculation as increasing the perturbative order of the inclusive charm production cross section does not necessarily translate into an increased accuracy of the entire dimuon process if the acceptance correction $\mathcal{A}$ is not updated likewise. Indeed, the acceptance correction $\mathcal{A}$ is typically calculated in a different framework such as the next-to-leading order (NLO) DISCO Monte Carlo program \cite{Kretzer:2001tc}. In this paper, we forgo the assumption in eq.~\eqref{eq:charm_production_factorization} and instead compute the dimuon cross section directly by implementing the entire chain of semi-inclusive charmed hadron production and the subsequent muonic decay. The calculation of semi-inclusive charmed hadron production effectively resums all logarithms $\log(Q^2/m^2_{\rm charm})$ that originate from collinear initial- and final-state QCD radiation into the scale dependence of PDFs and FFs. At this stage our implementation will be accurate at the next-to-leading order (NLO) level in perturbative QCD, accounting for the leading charm-mass corrections, though an extension to the full general-mass variable flavour number scheme (GM-VFNS) \cite{Kretzer:1997pd} will be straightforward. In addition, by using the recent next-to-NLO (NNLO) results for semi-inclusive DIS (SIDIS) publicly available from the literature \cite{Abele:2021nyo}, we are already able to estimate the expected size of NNLO corrections in our framework. The central idea is, that in the approach we present, one can systematically improve the theoretical description without a need for an external acceptance correction.

The paper is organized as follows. In section~\ref{sec:sidis}, we introduce the semi-inclusive calculation of charmed-hadron production in neutrino-nucleus DIS and in section~\ref{sec:decay} we discuss in detail the implementation of the charmed-hadron decay. In section~\ref{sec:results}, we then present our results for the dimuon cross sections comparing them with the NuTeV \cite{NuTeV:2007uwm} and CCFR \cite{CCFR:1994ikl} data, assess the uncertainties of our approach, and calculate the effective acceptance and nuclear corrections. Finally, section~\ref{sec:Conclusion} summarizes the work and discusses further developments.

% *****************************************************************************

\section{Semi-inclusive charged-current deep inelastic scattering}
\label{sec:sidis}

In the usual inclusive DIS processes, only the momentum of the outgoing lepton is measured and one integrates over the momenta of all other final-state particles. While experimentally simpler, this approach gives no information about the outgoing particles in the hadronic final state. In SIDIS, however, also the momentum of a particle in the hadronic final state is measured.

\begin{figure}[h!]
    \centering
    \includegraphics[]{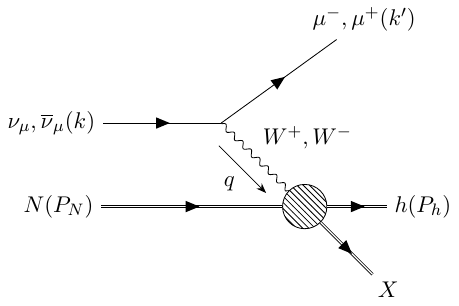}
    \caption{Diagrammatic depiction of the (anti)neutrino-SIDIS process $\nu_\mu N\to \mu hX$. Momenta of the particles are indicated in parentheses.}
    \label{fig:sidis}
\end{figure}

The charged-current SIDIS cross section for the production of a hadron $h$, depicted in figure~\ref{fig:sidis}, is given by
\begin{equation}
\label{eq:sidis_hadron_cross_section}
\begin{split}
	\frac{\dd\sigma(\nu_\mu N\to \mu h X)}{\dd x \, \dd y \, \dd z}&= \frac{G_F^2M_W^4}{\left(Q^2+M_W^2\right)^2}
  \frac{Q^2}{2\pi xy}   
 \\ &\phantom{=} \ \times \bigg[xy^2 F_1(x, z, Q^2)+\left(1-y-\frac{xy M^2}{s-M^2}\right)F_2(x, z, Q^2) \\ &\phantom{= \times\bigg[} \ \pm xy\left(1-\frac{y}{2}\right)F_3(x, z, Q^2)\bigg],
\end{split}
\end{equation}
where $G_F \approx 1.166379 \times 10^{-5} \, \mathrm{GeV}^{-2}$ is the Fermi coupling constant, $M_W\approx 80.377 \, \mathrm{GeV}$ is the mass of the $W^\pm$ boson, and the $+$ sign in the third term corresponds to neutrino scattering and the $-$ sign to antineutrino scattering. In eq.~\eqref{eq:sidis_hadron_cross_section} the usual DIS kinematical variables
\begin{equation}
\label{eq:sidis_kinematics}
\begin{aligned}
	Q^2&=-q^2 =- (k-k')^2 \,, \\
	x&=\frac{Q^2}{2P_N\cdot q} \,, \\
	y&=\frac{P_N\cdot q}{P_N\cdot k} \,,
\end{aligned}
\end{equation}
are accompanied by the momentum fraction of the produced hadron
\begin{equation}
\label{eq:sidis_kinematics_z}
	z=\frac{P_N\cdot P_h}{P_N\cdot q}.
\end{equation}
Here $P_N$, $P_h$, $k$, and $k'$ are the momenta of the target nucleon $N$, outgoing hadron $h$, incoming neutrino, and outgoing muon. At NLO, the zero-mass structure functions are given by
\begin{equation}
\label{eq:structure_functions}
\begin{aligned}
	F_2(x, z, Q^2)&=2x\sum_{ff'}\lambda_+^{ff'}\int_x^1 \frac{\dd\xi}{\xi}\int_z^1\frac{\dd\xi'}{\xi'} f(\xi, \fact^2)D_{f'\to h}(\xi', \frag^2) \\
		&\quad\quad \times\left[C_{\mathrm{LO}}^{ff'}(\hat x, \hat z)+\frac{\alpha_s(\renorm^2)}{2\pi}C_2^{ff'}(\hat x, \hat z; \fact^2, \frag^2)\right], \\
	F_1(x, z, Q^2)&=\sum_{ff'}\lambda_+^{ff'}\int_x^1 \frac{\dd\xi}{\xi}\int_z^1\frac{\dd\xi'}{\xi'} f(\xi, \fact^2)D_{f'\to h}(\xi', \frag^2) \\
		&\quad\quad \times\left[C_{\mathrm{LO}}^{ff'}(\hat x, \hat z)+\frac{\alpha_s(\renorm^2)}{2\pi}C_1^{ff'}(\hat x, \hat z; \fact^2, \frag^2)\right], \\
	F_3(x, z, Q^2)&=2\sum_{ff'}\lambda_-^{ff'}\int_x^1 \frac{\dd\xi}{\xi}\int_z^1\frac{\dd\xi'}{\xi'} f(\xi, \fact^2)D_{f'\to h}(\xi', \frag^2) \\
		&\quad\quad \times\left[C_{\mathrm{LO}}^{ff'}(\hat x, \hat z)+\frac{\alpha_s(\renorm^2)}{2\pi}C_3^{ff'}(\hat x, \hat z; \fact^2, \frag^2)\right],
\end{aligned}
\end{equation}
where $\hat x=x/\xi$ and $\hat z=z/\xi'$. The LO coefficient function is $C_{\mathrm{LO}}^{ff'}(\hat x, \hat z)=\delta(1-\hat x)\delta(1-\hat z)$ when $f$ and $f'$ are quarks, and zero otherwise. The couplings are given by
\begin{equation}
	\lambda_+^{ff'}=\begin{cases}
		\left|V_{ff'}\right|^2 &\text{$f$ and $f'$ are quarks} \\
		1 &\text{$f$ or $f'$ is a gluon}
	\end{cases},
\end{equation}
and
\begin{equation}
	\lambda_-^{ff'}=\begin{cases}
		-\lambda_+^{ff'} &\text{$f$ or $f'$ is an antiquark} \\
		\lambda_+^{ff'} &\text{otherwise}
	\end{cases}.
\end{equation}
The PDFs $f$ are evaluated at the factorization scale $\fact$, FFs $D_{f \rightarrow h}$ at the fragmentation scale $\frag$, and the strong coupling constant $\alpha_s$ at the renormalization scale $\renorm$. The quark flavors are determined by charge conservation. For the CKM matrix elements $V_{ff'}$, we use the global-fit values \cite{Workman:2022ynf}. The massless $\overline{\rm MS}$ NLO coefficient functions $C_i^{ff'}$ for $i=1,2,3$ can be found from refs.~\cite{Furmanski:1981cw} and \cite{deFlorian:1997zj}, where the latter also explicitly indicates the scale-dependent logarithms.\footnote{Note that there is a typographical error in ref.~\cite{Furmanski:1981cw}: on the second line of the coefficient $C_2^{F(1)}$ in Appendix II, the term $\frac{3(1+x)^2}{1-x}\ln x$ should be $\frac{3(1+x^2)}{1-x}\ln x$.} The NNLO coefficient functions are currently known only approximately. The massless quark-to-quark coefficient functions in the photon exchange are known in the next-to-leading power (NLP) approximation to NNLO \cite{Abele:2021nyo} and even to $\mathrm{N}^3\mathrm{LO}$ accuracy \cite{Abele:2022wuy}. Although these NNLO coefficient functions are approximate, we can still use them to get a rough estimate of the expected size of the NNLO corrections.\footnote{At the final stages of preparing this paper, preprints of the full massless NNLO coefficient functions for unpolarized and polarized photon-exchange SIDIS have appeared \cite{Goyal:2023xfi,Bonino:2024qbh,Bonino:2024wgg,Goyal:2024tmo}.} 
These corrections are given for photon-mediated interactions, so one has to adapt the electromagnetic couplings to the charged-current case. This is possible as the NLP terms are associated with emissions of soft gluons which do not affect the flavour structure of the process.

While quark masses are neglected in the hard coefficient functions $C_i^{ff'}$, we can take most of their effects into account by replacing $x$ in the right-hand side of eq.~\eqref{eq:structure_functions} by the so-called slow-rescaling variable \cite{PhysRevLett.36.1163,Barone:1995ma}
\begin{equation}
	\chi_q \equiv x\left(1+\frac{m_q^2}{Q^2}\right)>x \,, \label{eq:slowrescaling}
\end{equation}
where $m_q$ is the quark mass in channels where a single heavy quark is produced. If several heavy quarks are produced in the partonic scattering, this can be generalized. In practice, the only relevant channels turn out to be those in which a single charm quark is produced. The charm mass is taken to be that used in the PDFs. While the slow-rescaling variable induces an enhancement to $F_2$, as this structure function is directly proportional to $x$, the net effect also depends on the $x$-dependence of the PDF due to the shift in the momentum argument at which the PDFs are evaluated.

We note that introducing the rescaling variable $\chi_q$ is not just a prescription but it emerges naturally from the GM-VFNS calculations \cite{Nadolsky:2009ge}. Indeed, at leading order, the slow-rescaling variable actually captures all quark-mass effects. Thus any additional mass corrections are also suppressed by $\alpha_s$ and are, in general, of the order of $\mathcal{O}(\alpha_s m^2/Q^2)$. This is a specific feature of charged-current DIS, as there are additional mass effects already at the leading order in the neutral-current case. For a more quantitative check, we have computed the ratio of inclusive (i.e. not SIDIS) charm production with $m_c=1.5 \, {\rm GeV}$ and $m_c=1.3\, {\rm GeV}$. This ratio has been computed also in ref.~\cite{Gao:2017kkx}, including full quark-mass corrections in the fixed flavour-number scheme. Our calculation, with just the slow-rescaling variable, agrees to within ${5}{\, \%}$ with ref.~\cite{Gao:2017kkx} indicating that the rescaling variable already accounts for the majority of quark-mass effects and is an accurate-enough approach for our present purposes.

\section{Decay of charmed hadrons}
\label{sec:decay}

While the charmed hadrons cannot always be directly reconstructed by the experiments, their specific decay products can be measured. One of the prominent decay channels for charmed hadrons is $\mu+\text{anything}$, which produces the secondary decay muon in the dimuon production process. This process has been measured most recently by NuTeV \cite{NuTeV:2007uwm}, CCFR \cite{CCFR:1994ikl}, and NOMAD \cite{NOMAD:2013hbk}.

\begin{figure}[h!]
	\centering
    \includegraphics[]{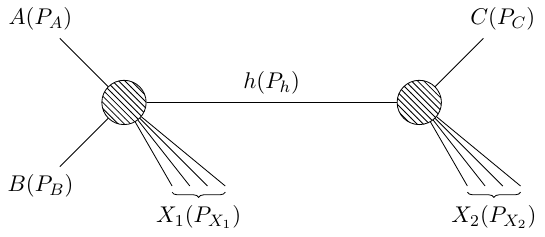}
	\caption{Diagrammatic depiction of the production process $AB\to CX$. Momenta of the particles are given in parentheses.}
	\label{fig:ABCX}
\end{figure}

\subsection{General formalism and the decay function}
\label{sec:formalism}

We begin by assuming that the cross section for the production of a charmed hadron $h$ in collisions between particles $A$ and $B$ with center-of-mass energy $\sqrt{s}$, followed by its decay to particle $C$, factorizes into a Breit-Wigner form,
\begin{equation}
\begin{split}
	\sigma(AB\to CX)=\frac{1}{2s}\int & \dd\Pi(P_C) \, \dd \Pi(P_{X_1}) \, \dd \Pi(P_{X_2}) \\ &\times (2\pi)^4\delta^{(4)}(P_A + P_B-P_C-P_{X_1}-P_{X_2}) \\ & \times \mathcal{A}_p \frac{1}{(P_h^2-m_h^2)^2+m_h^2 \Gammatot^2}\mathcal{A}_d \,,
\end{split}
\end{equation}
see figure~\ref{fig:ABCX}. Here, $\dd\Pi$ denote the phase space elements, $P_i$ are the momenta of different (groups of) particles, $m_h$ is the mass of the hadron $h$, and $\Gammatot$ denotes its total decay width. The symbols $\mathcal{A}_p$ and $\mathcal{A}_d$ denote the parts associated with the production and decay of the particle $h$, respectively. Here, we neglect the possible spin correlations between the production and decay, if the hadron $h$ is not a spin-$0$ particle. In the narrow-width approximation,
\begin{equation}
	\frac{1}{(P_h^2-m_h^2)^2+m_h^2\Gammatot^2}\simeq \frac{\pi}{m_h\Gammatot}\delta(P_h^2-m_h^2).
\end{equation}
In order to make the momentum $P_h$ of the intermediate-state particle explicit, we insert unity as
\begin{equation}
	1=\int\frac{\dd P_h^2}{2\pi}\int \frac{\dd^3\mathbf P_h}{(2\pi)^32E_h}(2\pi)^4\delta^{(4)}(P_h-P_C-P_{X_2}) \,,
\end{equation}
where $E_h$ is the energy of the hadron $h$. This results in
\begin{equation}
\begin{split}
	\sigma(AB\to CX)=\frac{1}{2s} & \int \frac{\dd^3\mathbf{P}_h}{(2\pi)^32E_h} \\
	& \times \int \dd\Pi(P_{X_1}) \, \mathcal{A}_p(2\pi)^4\delta^{(4)}(P_A+P_B-P_{X_1}-P_h) \\
	& \times \frac{1}{2m_h\Gammatot} \int \dd\Pi(P_{X_2}) \, \dd\Pi(P_C) \mathcal{A}_d(2\pi)^4\delta^{(4)}(P_h-P_C-P_{X_2}){\big|_{P_h^2=m_h^2}} \,.
\end{split}
\end{equation}
Identifying the differential production cross section with
\begin{equation}
	\frac{\dd\sigma(AB\to hX)}{\dd\Pi(P_h)}=\frac{1}{2s}\int\dd\Pi(P_{X_1}) \, \mathcal{A}_p (2\pi)^4\delta^{(4)}(P_A+P_B-P_{X_1}-P_h) \,,
\end{equation}
and the decay width of the hadron
\begin{equation}
\label{eq:decay_width}
	\Gamma_{h\to C}=\frac{1}{2m_h}\int \dd\Pi(P_{X_2}) \, \dd\Pi(P_C) \mathcal{A}_d(2\pi)^4\delta^{(4)}(P_h-P_C-P_{X_2}){\big|_{P_h^2=m_h^2}} \,,
\end{equation}
the total cross section becomes
\begin{equation}
\label{eq:total_decay_cross_section}
	\sigma(AB\to CX)=\int \dd\Pi(P_h)\frac{\dd\sigma(AB\to h X)}{\dd\Pi(P_h)}\frac{\Gamma_{h\to C}}{\Gammatot} \,.
\end{equation}

We now want to introduce a kinematical cut on the energy of the outgoing particle $C$, ${E_C \geq E_C^{\mathrm{min}}}$. This is needed for dimuon production, as the NuTeV and CCFR data include a cut on the decay muon energy. To take such a cut into account, we have to make eq.~\eqref{eq:total_decay_cross_section} differential in the momentum of $C$. This amounts to replacing the decay width $\Gamma_{h\to C}$ with a more differential quantity. To this end, we define a dimensionless decay function $d_{h\to C}$,
\begin{equation}
\label{eq:d_definition}
	d_{h\to C}(w)=\frac{1}{2(2\pi)^3}\int \dd\Pi(P_{X_2}) \mathcal{A}_d (2\pi)^4\delta^{(4)}(P_h-P_C-P_{X_2}){\big|_{P_h^2=m_h^2}} \,,
\end{equation}
where
\begin{align}
w \equiv \frac{P_C\cdot P_h}{m_h^2}  \,, \quad 0 \leq w \leq \frac{1}{2} \,.
\end{align}
To see why $d_{h\to C}$ depends only on $w$, note that $d_{h\to C}$ is a Lorentz scalar depending on the momenta $P_h$ and $P_C$ and thus the only non-trivial invariant is $P_C\cdot P_h$. As the mass of $C$ will be negligible in the present case, it makes sense to normalize this with $P_h^2=m_h^2$. Using eq.~\eqref{eq:d_definition}, the decay width in eq.~\eqref{eq:decay_width} can be written as
\begin{equation}
	\Gamma_{h\to C}=\frac{1}{2m_h}\int \frac{\dd^3\mathbf P_C}{E_C} d_{h\to C}(w)=\frac{1}{2m_h}\int \dd|\mathbf P_C| \, \dd\Omega \, \frac{|\mathbf P_C|^2}{E_C} d_{h\to C}(w) \,.
\end{equation}
The differential decay width is then
\begin{equation}
\label{eq:differential_decay_width}
	\frac{\dd\Gamma_{h\to C}}{\dd|\mathbf P_C|}=\frac{\pi}{m_h}\frac{|\mathbf P_C|^2}{E_C}\int \dd(\cos\theta) \, d_{h\to C}(w) \,,
\end{equation}
where the angular dependence in $w$ is given explicitly by
\begin{equation}
\label{eq:w_argument}
	w=\frac{1}{m_h^2}\left(E_C E_h-|\mathbf P_C||\mathbf P_h|\cos\theta\right) \,,
\end{equation}
and $\theta$ denotes the relative angle between ${\bf P}_C$ and ${\bf P}_h$. Integrating eq.~\eqref{eq:differential_decay_width} over $|\mathbf P_C|$ and inserting it back to eq.~\eqref{eq:total_decay_cross_section},
\begin{equation}
	\dd\sigma(AB\to CX)=\frac{\pi}{m_h \Gammatot}\int \dd\Pi(P_h) \, \frac{\dd\sigma(AB\to hX)}{\dd\Pi(P_h)}\int \dd|\mathbf P_C| \, \frac{|\mathbf P_C|^2}{E_C}\int \dd(\cos\theta) \, d_{h\to C}(w) \,.
\end{equation}
For convenience, we define $\rho \equiv E_C/E_h$ and, neglecting the mass of $C$, introduce an energy-dependent branching fraction $B_{h\to C}(E_h, E_C^{\mathrm{min}})$ as
\begin{equation}
\label{eq:B_definition}
	B_{h\to C}(E_h, E_C^{\mathrm{min}}) \equiv \frac{\pi}{m_h \Gammatot}\int \dd\rho \, \rho E_h^2\int \dd(\cos\theta) \, d_{h\to C}(w){\big|_{E_C=\rho E_h\geq E_C^{\mathrm{min}}}} \,,
\end{equation}
which allows us to write the cross section simply as
\begin{equation}
\label{eq:sidis_cross_section}
	\dd\sigma(AB\to CX) = \int \dd \Pi(P_h) \, \frac{\dd\sigma(AB\to hX)}{\Pi(P_h)} B_{h\to C}(E_h, E_C^{\mathrm{min}}).
\end{equation}
Combining eqs.~\eqref{eq:differential_decay_width} and \eqref{eq:B_definition}, we see that
\begin{equation}
	B_{h\to C}(E_h, E_C^{\mathrm{min}})=\frac{1}{\Gammatot}\int\dd^3\mathbf P_C\frac{\dd\Gamma_{h\to C}}{\dd^3\mathbf P_C}\bigg|_{E_C\geq E_C^{\mathrm{min}}}.
\end{equation}
When there is no cut, $E_C^{\mathrm{min}}=0$, $B_{h\to C}$ becomes independent of $E_h$ and reduces to the usual branching fraction $\Gamma_{h\to C}/\Gammatot$. For a non-zero energy cut, $B_{h\to C}$ depends on the energy $E_h$. Its dependence on $E_h$ can be precomputed in a dense-enough grid, such that the double integrals in eq.~\eqref{eq:B_definition} do not need to be evaluated repeatedly. This significantly speeds up the evaluation of the cross section in eq.~\eqref{eq:sidis_cross_section}.

We will apply the formalism introduced above to the case of dimuon production process $\nu_\mu N\to \mu^-\mu^+ X$. Using eq.~\eqref{eq:sidis_cross_section}, we can write the production cross section as
\begin{equation}
	\frac{\dd\sigma(\nu_\mu N\to \mu^- \mu^+ X)}{\dd x \, \dd y}=\sum_h \int_{z_{\text{min}}}^1 \dd z \, \frac{\dd\sigma(\nu_\mu N\to \mu^- h X)}{\dd z \, \dd x \, \dd y}B_{h\to \mu}(E_h=zyE_\nu, E_{\mu}^{\text{min}}),
\label{eq:SIDISdecay}
\end{equation}
where the sum is over $h\in\left\{D^0, D^+, D_s, \Lambda^+_c\right\}$. The kinematical variables $x$, $y$, and $z$ were defined in eqs.~\eqref{eq:sidis_kinematics} and \eqref{eq:sidis_kinematics_z}. The lower bound of integration $z_{\mathrm{min}}$ is given by
\begin{equation}
	z=\frac{E_h}{yE_\nu}\geq \frac{\max\{ E_{\mu}^{\mathrm{min}}, m_h\}}{y E_\nu}\equiv z_{\text{min}} \,,
\end{equation}
where the relation $E_h=zyE_\nu$ holds in the rest frame of the nuclear target.

\subsection{Fitting the decay function}
\label{sec:decay_function_fit}

For the dimuon production process, we need to determine $d_{h\to\mu}$. To do this, we adapt the formalism introduced in the previous section to $D$-meson production in $e^+e^-$ collisions. We use data from CLEO \cite{CLEO:2006ivk} given for the semileptonic decays $D\to e^+\nu_eX$. While this technically corresponds to $C=e^+$ and not $C=\mu^+$, we assume that the differences are negligible as both the electron and muon masses are small compared to the $D$-meson masses. 

The CLEO data are collected at the $\psi(3770)$ mass threshold. The $\psi(3770)$ decays mostly (with a branching fraction of $\left(93 \substack{+8 \\ -9} \right)\,\%$ \cite{Workman:2022ynf}) into a $D$-meson pair $D\overline D$, where the relevant $D^0$ and $D^+$ mesons are not produced to rest. Instead, they carry momenta ${0.277}{\, {\rm GeV}}$ and ${0.243}{\, {\rm GeV}}$ for the $D^0$ and $D^+$ meson, respectively. Although the CLEO data are given separately for the $D^0$ and $D^+$ mesons, we combine them into one universal decay function that we use for all charmed hadrons. This is a reasonable assumption since the decay function depends only on $w$, which by eq.~\eqref{eq:w_argument} depends on $m_h$ and $\left|\mathbf P_h\right|$. Both of these values are very similar for $D^0$ and $D^+$.

\begin{figure}[b!]
    \centering
	\includegraphics[width=0.5\textwidth]{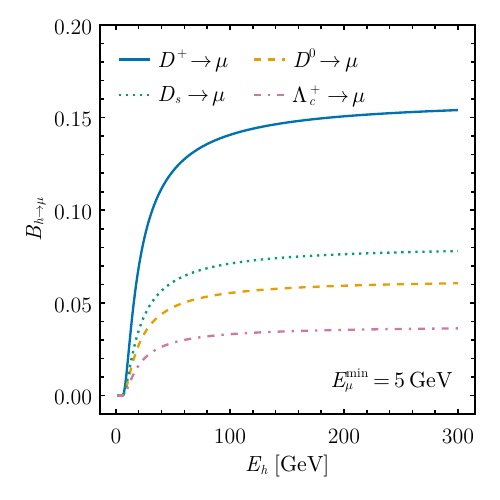}
	\caption{The branching fractions $B_{h\to\mu}$ of eq.~\eqref{eq:B_definition} as a function of the hadron energy $E_h$ with $E_{\mu}^{\mathrm{min}}={5}{\, {\rm GeV}}$ for $D^+$, $D^0$, $D_s$, and $\Lambda^+_c$. For the decay function, we use eq.~\eqref{eq:decay_function_parametrization} and parameters given in eq.~\eqref{eq:decay_best_fit_params}.}
	\label{fig:branching_fraction_plots}
\end{figure}

The data are fitted with the parametrization
\begin{equation}
	\label{eq:decay_function_parametrization}
	d_{h\to \mu}(w)=Nw^\alpha(1-\gamma w)^\beta \,,
\end{equation}
for $0\leq w \leq 1/\gamma$ and zero elsewhere. The parameter $\gamma$ is expected to be around 2, since $w \leq 1/2$. For the fit, we combine the two CLEO data sets for $D^0$ and $D^+$, and take the $D$-meson mass and momentum parameters to be the mean values of the corresponding $D^0$ and $D^+$ values: $m_h={1.867}{\, {\rm GeV}}$ and $|\mathbf P_h|={0.26}{\, {\rm GeV}}$. The best-fit parameters are then
\begin{equation}
\label{eq:decay_best_fit_params}
	N=3.156, \quad \alpha = 1.009, \quad \beta = 1.799, \quad \gamma = 2.101,
\end{equation}
which are obtained by the nonlinear least-squares method. Figure~\ref{fig:branching_fraction_plots} shows the branching fractions $B_{h\to\mu}$ of eq.~\eqref{eq:B_definition} using the parametrization eq.~\eqref{eq:decay_function_parametrization} and parameter values in eq.~\eqref{eq:decay_best_fit_params}. Here, and in the rest of the paper, we have assumed that the same decay function $d_{h\to \mu}$ applies for all charmed hadrons. The asymptotic values as $E_h\to\infty$ are consistent with the usual branching fractions \cite{Workman:2022ynf} which gives us confidence concerning this universality assumption of $d_{h\to \mu}$. For example, the semimuonic branching ratio of $D^+$ is $17.6 \pm 3.2 \, \%$ \cite{Workman:2022ynf}, which is in line with figure~\ref{fig:branching_fraction_plots}.

\begin{figure}[hb!]
\centering
	\includegraphics[width=0.49\linewidth]{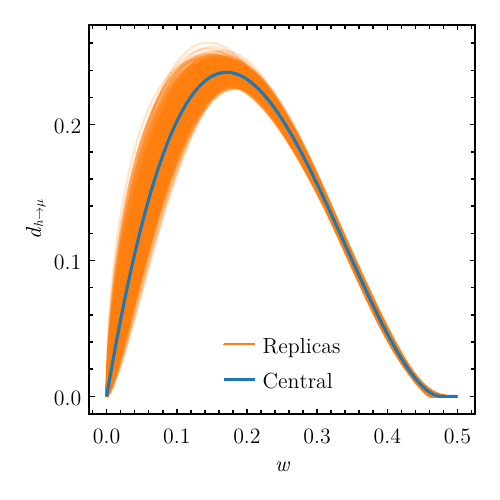}
	\includegraphics[width=0.49\linewidth]{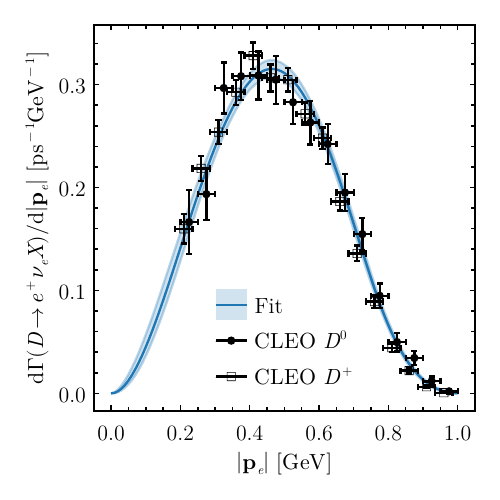}
	\caption{{\bf Left:}
	The decay function $d_{h\to\mu}$ of eq.~\eqref{eq:decay_function_parametrization} as a function of $w$. The central curve corresponds to the best-fit parameters given in eq.~\eqref{eq:decay_best_fit_params}, while the replica curves correspond to all of the replica fit parameters. {\bf Right:}
	Differential decay width of $D\to e^+\nu_e X$ as a function of the $D$ meson momentum, with $D^0$ and $D^+$ data from ref.~\cite{CLEO:2006ivk}. The $D^+$ data points have been slightly shifted horizontally for improved readibility. The central fit curve corresponds to eq.~\eqref{eq:differential_decay_width} with the parameter values given in eq.~\eqref{eq:decay_best_fit_params}. The band depicts the ${90}{\, \%}$ confidence-level uncertainty taken from the replica fits.}
	\label{fig:decay_model_envelope_decay_parametrization}
\end{figure}

The fit parameters are highly correlated. To assess the uncertainty of the fit parameters, we employ a strategy similar to that of refs.~\cite{Forte:2002fg,Ball:2008by}. From the original CLEO data set, we generate 1000 new replica data sets so that each data value $D_i$ is transformed into
\begin{equation*}
	D_i'=D_i(1+\delta D_i R_i) \,,
\end{equation*}
where $R_i$ is a random number drawn from the standard normal distribution and $\delta D_i$ is the total relative uncertainty of the data point.\footnote{A more appropriate choice would be to only include statistical uncertainty, but the CLEO data includes only the total uncertainty.} We then fit these replica data sets with the same procedure as above to obtain a set of replica fits. An observable can then be evaluated using all replica fits to estimate the uncertainty resulting from the fitting procedure. In order to combine the results obtained with the replica fits into an uncertainty band, we can either take some central subset of all replica results as in ref.~\cite{AbdulKhalek:2022fyi} or simply compute the standard deviation. We have found that in the present case the differences between these two approaches are negligible. The left-hand panel of figure~\ref{fig:decay_model_envelope_decay_parametrization} shows the central result for $d_{h\to\mu}$ accompanied by the 1000 replica fits, and the right-hand panel compares the fit and its 90\,\% uncertainty with the CLEO data. This $90\,\%$ uncertainty band is derived by rejecting $5\,\%$ of the upper- and lower-extreme replica sets.

% *************************************************************

\section{Results}
\label{sec:results}

Having the decay function for $D$ mesons fitted to the CLEO data we can use it in eq.~\eqref{eq:SIDISdecay} and calculate dimuon production cross sections in neutrino-nucleus SIDIS without resorting to the approximate factorization of the acceptance correction discussed in the context of eq.~\eqref{eq:charm_production_factorization}. We will compare these calculations with data from the NuTeV \cite{NuTeV:2007uwm} and CCFR \cite{CCFR:1994ikl} experiments. These data correspond to $E_{\mu}^{\rm min} = 5\,{\rm GeV}$. Similar dimuon data exist also from the NOMAD collaboration \cite{NOMAD:2013hbk}. In this case the data are provided as a ratio between the fully integrated dimuon and inclusive DIS processes which is beneficial in terms of cancellation of experimental systematic uncertainties. However, the minimum allowed $Q^2$ is very low, $Q^2 > 1\,{\rm GeV}^2$, and the observable is sensitive to the region which begins to be outside the regime where our calculation can be applied. In addition, an acceptance correction has already been applied to account for the finite experimental cut in the final-state muon energy $E_{\mu}^{\rm min}$, whereas the entire point of the present paper is to avoid such external correction. While we could technically set $E_{\mu}^{\rm min}=0$, this would make us also increasingly sensitive to the small-$z$ instabilities of the time-like $Q^2$ evolution of the FFs, and also to the region of $D \rightarrow e^+X$ transition not directly constrained by the CLEO data. As a result, we only consider here the case with a finite $E_{\mu}^{\rm min}$.

For the nuclear PDFs, we use the \texttt{EPPS21} \cite{Eskola:2021nhw}, \texttt{nCTEQ15HQ} \cite{Duwentaster:2022kpv}, and \texttt{nNNPDF3.0} \cite{AbdulKhalek:2022fyi} NLO sets. The charm mass, which enters the cross-section calculation through the slow-rescaling variable, is taken to be the same as in the PDF set, i.e. $\SI{1.3}{GeV}$ for \texttt{EPPS21} and \texttt{nCTEQ15HQ}, and $\SI{1.51}{GeV}$ for \texttt{nNNPDF3.0}. For the fragmentation functions, we use the GM-VFNS NLO set \texttt{kkks08} \cite{Kneesch:2007ey} for $D^0$ and $D^+$, and NLO \texttt{bkk05} \cite{Kniehl:2006mw} for $D_s$ and $\Lambda_c^+$. Both sets use a value of $\SI{1.5}{GeV}$ for the charm-quark mass, which is also the parametrization scale in these analyses. In the case of \texttt{kkks08}, the provided sets include a fit using $e^+ e^-$ collision data only from the OPAL/LEP \cite{OPAL:1996ikk,OPAL:1997edj} experiment at $\sqrt{s} \approx m_Z\approx \SI{91}{GeV}$, one with data only from the Belle \cite{Belle:2005mtx} experiment at $\sqrt{s} \approx \SI{10}{GeV}$, and a global fit including data from several experiments. As was pointed out in ref.~\cite{Kneesch:2007ey}, some tension was found between the low- and high-energy data sets, which did not allow for a satisfactory fit when combining these different data sets. In this study we choose to use only the OPAL data, as this should be theoretically cleanest as potential ambiguities due to mass effects ought to be negligible at these high energies. 

\subsection{Uncertainties from the decay function}

\begin{figure}[t!]
\centering
	\begin{minipage}{0.48\textwidth}
		\includegraphics[width=\linewidth]{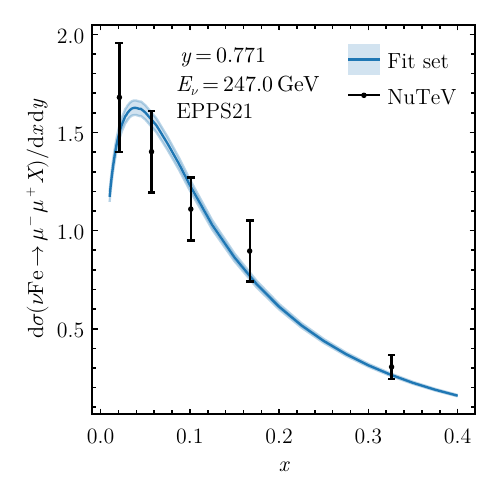}
	\end{minipage}%
	\quad%
	\begin{minipage}{0.48\textwidth}
		\includegraphics[width=\linewidth]{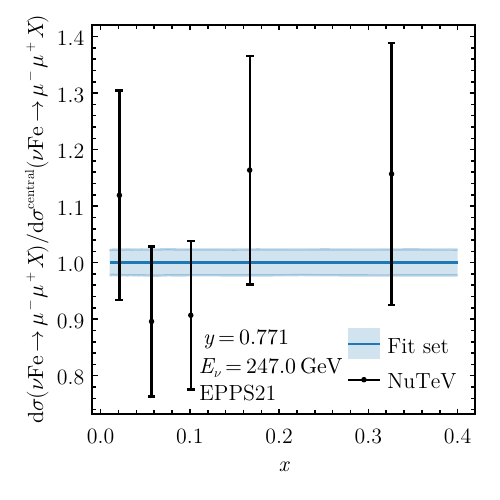}
	\end{minipage}
	\caption{\textbf{Left:} Neutrino dimuon cross section evaluated using \texttt{EPPS21} \cite{Eskola:2021nhw} at $y=0.771$ and $E_\nu={247}{\, {\rm GeV}}$. The central curve uses the parameter values given in eq.~\eqref{eq:decay_best_fit_params}. The band depicts the ${90}{\, \%}$ confidence-level uncertainty taken from the replica fits. The theoretical computation is compared against NuTeV data from ref.~\cite{NuTeV:2007uwm}. The cross-section values should be multiplied by $G_F^2 ME_\nu/100\pi$. \textbf{Right:} Same as on the left, but normalized to the central result with \texttt{EPPS21}.}
	\label{fig:decay_parametrization_effects}
\end{figure}

We begin by quantifying the uncertainties associated with the decay function described in section~\ref{sec:decay_function_fit}. Figure~\ref{fig:decay_parametrization_effects} indicates that the cross sections are very stable in terms of variations of the parameters and the uncertainty associated with the fitting procedure is minimal with a relative uncertainty of ${2}{\, \%}$, which is of the same order as the uncertainty in the right-hand panel of figure~\ref{fig:decay_model_envelope_decay_parametrization}.

\subsection{Channel decomposition}

\begin{figure}[p!]
	\begin{subfigure}{0.495\textwidth}
		\includegraphics[width=\linewidth]{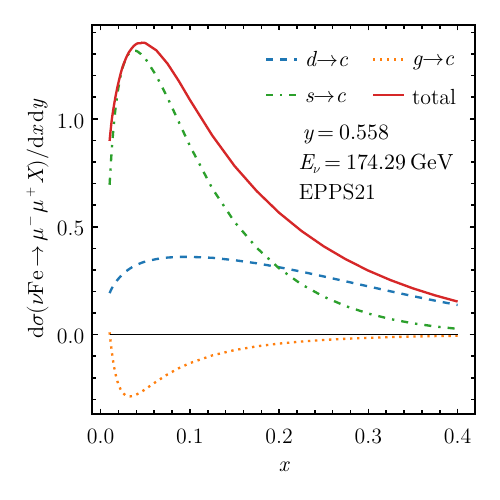}
	\end{subfigure}
	\begin{subfigure}{0.495\textwidth}
		\includegraphics[width=\linewidth]{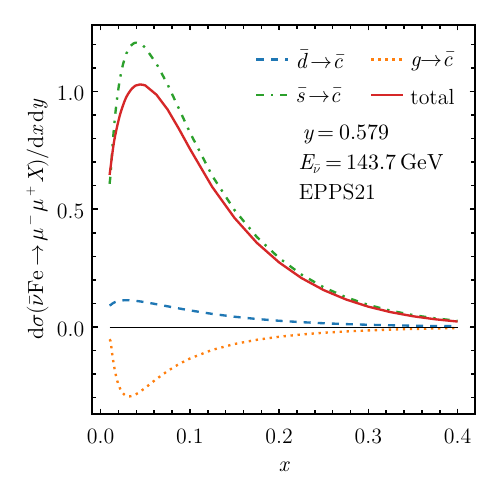}
	\end{subfigure}
	\caption{
 Contributions of different partonic channels to the neutrino (left) and antineutrino (right) dimuon cross sections. Only the relevant channels are shown, as the other channels are negligible. The cross section is computed with \texttt{EPPS21} \cite{Eskola:2021nhw} in the NuTeV kinematics. The cross-section values should be multiplied by $G_F^2 ME_\nu/100\pi$.}
	\label{fig:flavor_decomposition}
\end{figure}

\begin{figure}[p!]
	\centering
    \begin{subfigure}{0.495\textwidth}
		\includegraphics[width=\linewidth]{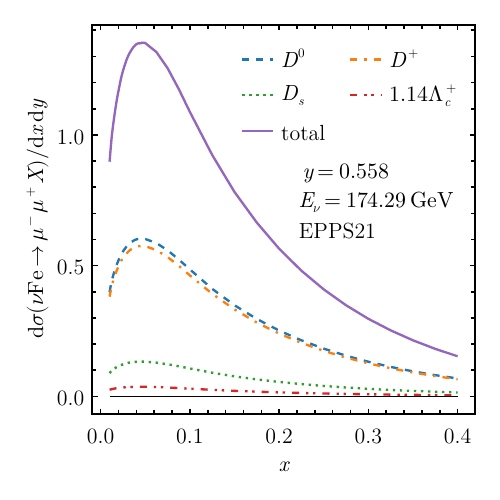}
	\end{subfigure}%
	\begin{subfigure}{0.495\textwidth}
		\includegraphics[width=\linewidth]{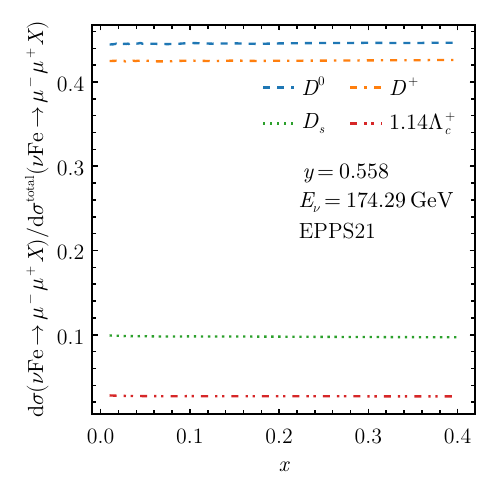}
	\end{subfigure}
	\caption{\textbf{Left:} 
      Contributions of different charmed hadrons to the neutrino dimuon cross section, computed with \texttt{EPPS21} \cite{Eskola:2021nhw} at $y=0.558$ and $E_\nu=\SI{174.29}{GeV}$. The sum of the individual contributions is shown with the solid curve. The contribution of $\Lambda_c^+$ is multiplied by $1.14$. The cross-section values should be multiplied by $G_F^2 ME_\nu/100\pi$. \textbf{Right:} The same as on the left, but as ratios to the total cross section.}
	\label{fig:fragmentation_decomposition}
\end{figure}

Next, we study the sensitivity of the cross sections to different partonic production channels by plotting some contributions separately in figure~\ref{fig:flavor_decomposition}. Only those contributions in which there is a final-state charm quark present are shown, as the contributions from gluons and other quarks fragmenting into a charmed hadron turned out negligible. As expected from the CKM matrix, the most relevant channel is the $s\to c$ channel ($\bar s\to \bar c$ for antineutrino scattering), where a strange quark is taken from the target and a charm quark fragments into a charmed hadron. In the case of neutrino scattering, the $d$-quark contribution becomes dominant at $x\gtrsim 0.2$. This behavior reflects the fact that the $d$-quark distribution includes a valence contribution that increases the PDF at large values of $x$ and compensates for the smaller CKM matrix element. In comparison, the $\bar{d}$-quark initiated processes in the antineutrino scattering have only a small contribution to the cross section as the $\bar{d}$-quark PDFs are of the same order as for $\bar{s}$ quarks. Interestingly, we notice that at NLO the gluon-initiated processes have a negative contribution to the cross section, which is related to the definitions of the NLO PDFs and FFs in the $\msbar$ scheme \cite{Candido:2020yat}. Overall, these results confirm that the considered cross sections are very sensitive to the $s$- and $\bar{s}$-quark PDFs.

The contributions of different charmed hadrons to the full dimuon cross section are shown in figure~\ref{fig:fragmentation_decomposition}. We find that the relative contributions from different hadrons are independent of the kinematics as the relative contributions reflect the ratio
\begin{equation}
	\frac{f(c\to h)/(m_h \Gamma_h^{\text{tot}})}{\sum_h f(c\to h)/(m_h \Gamma_h^{\text{tot}})} \,,
 \label{eq:frag_fractions}
\end{equation}
which can be understood from eqs.~\eqref{eq:B_definition} and \eqref{eq:SIDISdecay}. The fragmentation fractions $f(c\to h)$ for each hadron $h$ can be found e.g. in refs.~\cite{ZEUS:2005pvv,Lisovyi:2015uqa} and the total decay widths $\Gamma_h^{\text{tot}}$ in ref.~\cite{Workman:2022ynf}. The contribution of $\Lambda_c^+$ is multiplied by $1.14$ to account for other baryonic states \cite{ZEUS:2005pvv}. The biggest contributions of roughly $\SI{40}{\percent}$ are from $D^0$ and $D^+$ production, with $D_s$ contributing around $\SI{10}{\percent}$. The smallest contribution comes from $\Lambda_c^+$ (and other baryon states) at about $\SI{3}{\percent}$. The calculated fractions are indeed well in line with the expectations of eq.~\eqref{eq:frag_fractions}.

\subsection{Scale dependence}
\label{sec:uncertainty_estimation}

To gauge the effects of missing higher orders, we employ the standard approach of computing observables with different combinations of the renormalization scale $\renorm$, factorization scale $\fact$, and fragmentation scale $\frag$. These scales are varied for the combinations $\renorm, \fact, \frag \in \{\frac{1}{2}Q, Q, 2Q\}$, with the restriction $\frac{1}{2}\renorm \leq \fact, \frag \leq 2\renorm$ to limit the appearance of excessively large scale logarithms. This results in 17 different scale-choice combinations from which the resulting error band is derived by taking an envelope. For some NuTeV and CCFR data points, the scale at which the PDFs and FFs are evaluated becomes smaller than the parametrization scale of the respective analyses.
This is especially the case when computing the scale dependence with $\fact, \frag=\frac{1}{2}Q$. In these cases, we freeze the factorization scales to the minimum values of $\mu_{\rm fact}$ specified by the PDF analysis and do not extrapolate beyond the edge of the grid. As we use the values for $\alpha_s$ from the PDF sets, we freeze also the renormalization scale $\renorm$ to the same value as $\fact$. Below the charm mass-threshold $m_c=\SI{1.5}{GeV}$, the charm fragmentation functions are scale-independent, see e.g. the discussion in ref.~\cite{Epele:2016gup}.

\begin{figure}
    \centering
	\begin{subfigure}{0.49\textwidth}
		\includegraphics[width=\linewidth]{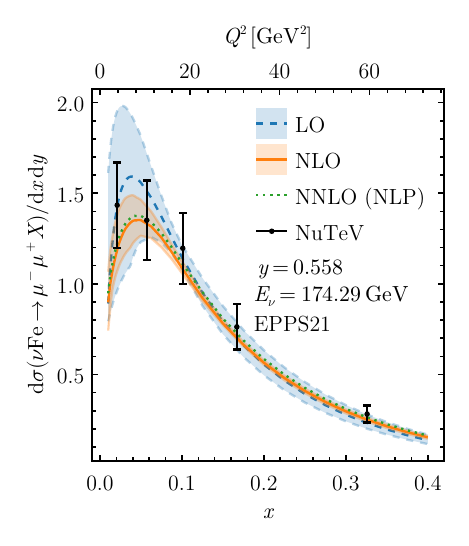}
	\end{subfigure}
	\begin{subfigure}{0.49\textwidth}
		\includegraphics[width=\linewidth]{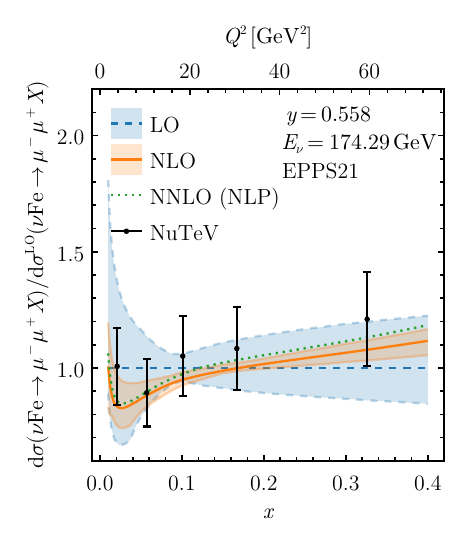}
	\end{subfigure}
	\begin{subfigure}{0.49\textwidth}
		\includegraphics[width=\linewidth]{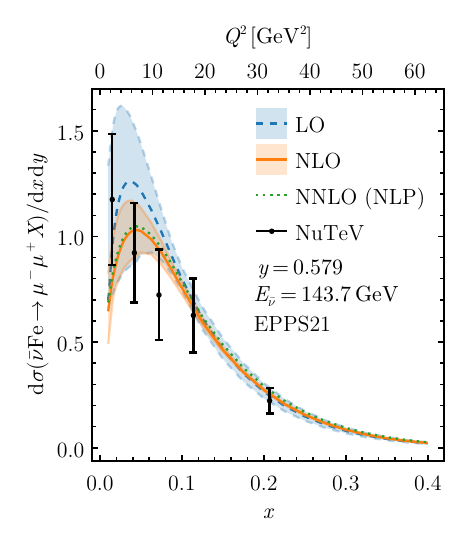}
	\end{subfigure}
	\begin{subfigure}{0.49\textwidth}
		\includegraphics[width=\linewidth]{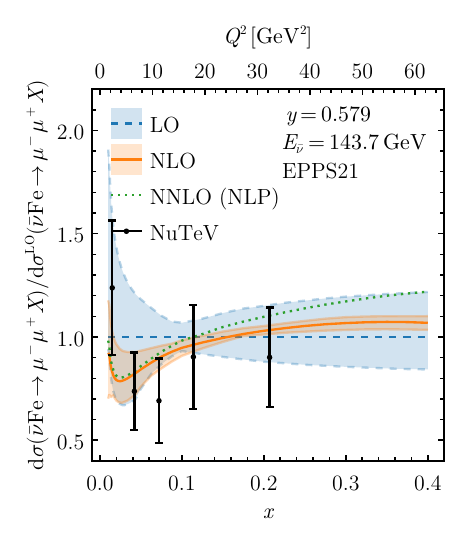}
	\end{subfigure}
    \caption{Scale uncertainties in the neutrino (top) and antineutrino (bottom) dimuon cross sections, computed with \texttt{EPPS21} \cite{Eskola:2021nhw}. Absolute cross sections are shown on the left and cross sections normalized to the central LO values on the right. The perturbative order in the figure refers only to the partonic cross section. In each case, the PDF is evaluated at NLO. The NNLO (NLP) refers to the approximate NNLO calculation of ref. \cite{Abele:2021nyo}. Theoretical predictions are compared with the NuTeV data \cite{NuTeV:2007uwm}. The uncertainty band depicts the envelope of all scale choice combinations. The cross-section values should be multiplied by $G_F^2 ME_\nu/100\pi$.}
    \label{fig:scale_dependence}
\end{figure}

The resulting uncertainties from the scale variations for the dimuon production cross section in neutrino and antineutrino scattering are shown in figure~\ref{fig:scale_dependence} for LO and NLO, including also the central, approximate NNLO correction. We have not computed an uncertainty band associated with the approximate NNLO coefficient functions. For this comparison, figure~\ref{fig:scale_dependence} shows only representative kinematic bins of $y$ and $E$. As is to be expected, moving from the LO to the NLO calculation decreases the uncertainties from the scale variations significantly. The scale uncertainty also grows larger at smaller $x$, which corresponds to smaller $Q^2$. Towards this region, any calculation based on perturbative QCD becomes less reliable due to the stronger scale dependence of $\alpha_s$ and PDFs. In addition, effects from the missing mass terms at the matrix-element level could become more relevant. It is also possible that higher-twist effects start to play a role when $Q^2$ is small enough. 

The NLO corrections are rather significant throughout the available kinematic range but stay below $\sim 30\,\%$ apart from the lowest values of $Q^2$. 
At small values of $x$, the approximate NNLO corrections are typically small. Towards larger values of $x$ their relevance appears to increase and in the case of neutrino scattering the approximate NNLO result tends to lie near the upper edge of the NLO scale-uncertainty band. In the antineutrino-scattering case, the relative NNLO correction is larger than in the neutrino-scattering case due to the different $x$-dependence of the $\bar{d}$-quark distribution compared to the $d$-quark distribution.
However, we stress that here we have applied only the approximative NNLO coefficient functions and the situation may still change once the full charged-current NNLO coefficient functions are available.

\subsection{Nuclear-PDF uncertainties}

\begin{figure}[p!]
	\centering
	\includegraphics[width=\linewidth]{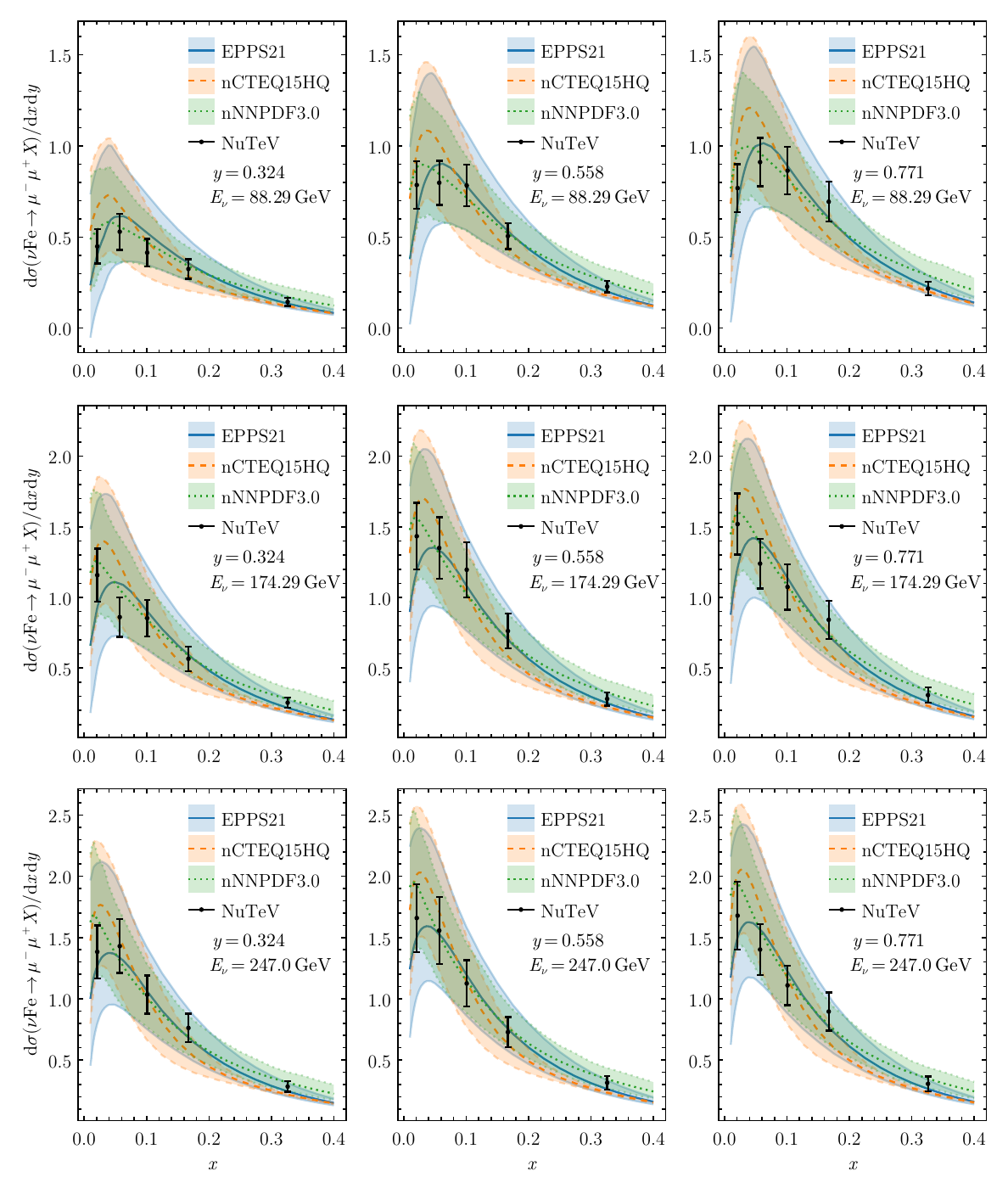}
	\caption{Neutrino dimuon cross sections in the NuTeV kinematics evaluated using the \texttt{EPPS21} \cite{Eskola:2021nhw}, \texttt{nCTEQ15HQ} \cite{Duwentaster:2022kpv}, and \texttt{nNNPDF3.0} \cite{AbdulKhalek:2022fyi} PDF sets. The calculations are done at next-to-leading order. The uncertainty bands depict the PDF uncertainties with a $\SI{90}{\percent}$ confidence interval. The theoretical calculations are compared against NuTeV data from ref.~\cite{NuTeV:2007uwm}. The cross-section values should be multiplied by $G_F^2 ME_\nu/100\pi$.}
	\label{fig:nutev_neutrino_pdf_comparison}
\end{figure}

\begin{figure}[p!]
	\centering
	\includegraphics[width=\linewidth]{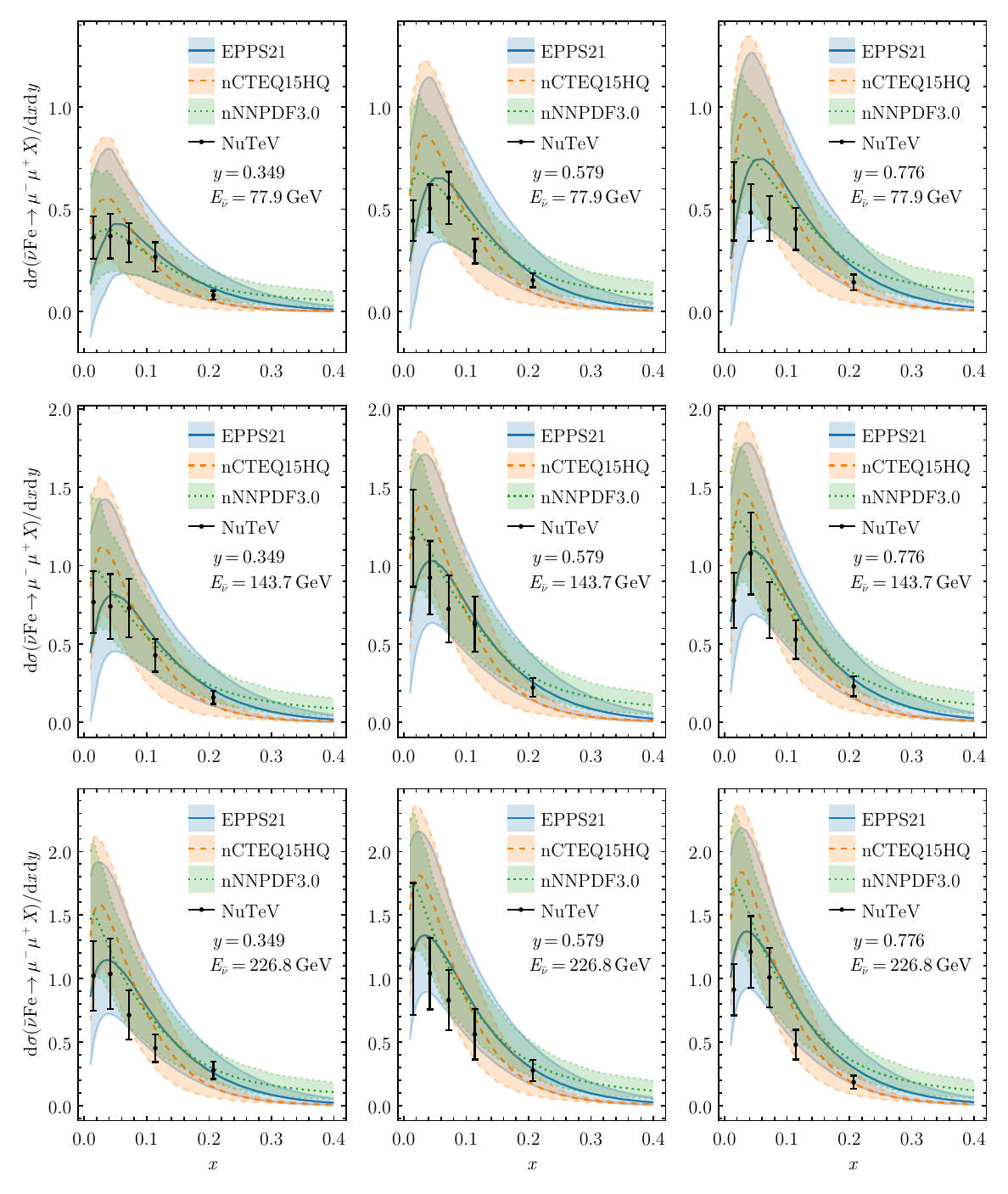}
	\caption{Same as figure~\ref{fig:nutev_neutrino_pdf_comparison}, but for antineutrino scattering.}
	\label{fig:nutev_antineutrino_pdf_comparison}
\end{figure}

\begin{figure}[p!]
	\begin{subfigure}{0.495\textwidth}
		\includegraphics[width=\linewidth]{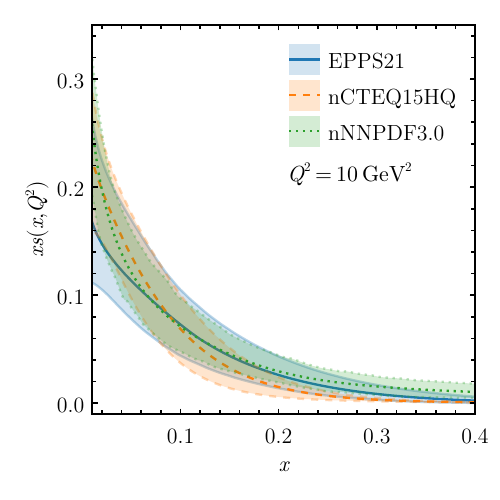}
	\end{subfigure}
	\begin{subfigure}{0.495\textwidth}
		\includegraphics[width=\linewidth]{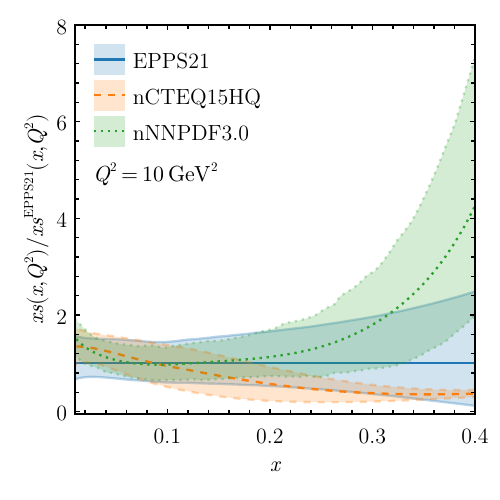}
	\end{subfigure}
	\caption{\textbf{Left:} The strange-quark distribution $xs(x, Q^2=\SI{10}{GeV^2})$ for the \texttt{EPPS21} \cite{Eskola:2021nhw}, \texttt{nCTEQ15HQ} \cite{Duwentaster:2022kpv}, and \texttt{nNNPDF3.0} \cite{AbdulKhalek:2022fyi} nuclear PDFs. The band indicates the $\SI{90}{\percent}$ confidence-level uncertainties of the nuclear PDFs. \textbf{Right:} The same distribution as on the left, but normalized to the central \texttt{EPPS21} value.}
	\label{fig:npdf_xs_10}
\end{figure}

We will now compare our NLO calculations with different nuclear PDFs with the ex\-perimental data from NuTeV and CCFR.
Figures~\ref{fig:nutev_neutrino_pdf_comparison} and \ref{fig:nutev_antineutrino_pdf_comparison} compare the NuTeV \cite{NuTeV:2007uwm} cross sections on dimuon production in neutrino-nucleus and antineutrino-nucleus scatterings with NLO calculations using \texttt{EPPS21}, \texttt{nCTEQ15HQ}, and \texttt{nNNPDF3.0} nuclear PDFs. While the calculations with the three different nuclear PDF sets are in agreement with each other when their uncertainties are accounted for, some systematic differences can be observed. 
At the smallest values of $x$, the \texttt{EPPS21} calculation gives a lower cross section than the other sets, whereas at $0.1<x<0.4$, \texttt{nCTEQ15HQ} is below the others. For $x>0.2$ the \texttt{nNNPDF3.0} set, in turn, yields the highest cross section. These differences directly reflect the differences in the general shapes of the strange-quark distributions, as is seen in figure~\ref{fig:npdf_xs_10}, further underlying the sensitivity of this observable to the $s$-quark PDF. The calculations with all three nuclear PDFs describe the NuTeV data well within the uncertainties, though at $x<0.1$ the data seem to slightly prefer the lower cross section from \texttt{EPPS21} compared to the others. We note here that these data were included in the fit of \texttt{nNNPDF3.0} nuclear PDFs, where the factorized acceptance correction was used. In the case of antineutrino scattering, the cross sections are slightly smaller overall and the fall-off as $x$ increases is steeper. This is related to having a significant large-$x$ contribution from the valence $d$-quark PDFs in the case of neutrino scattering, as was shown in figure~\ref{fig:flavor_decomposition}. In general, the systematics of the relative behaviour of calculations with different nuclear PDFs are very similar between neutrino and antineutrino scattering and within uncertainties all sets describe the data. 

\begin{figure}[p!]
	\begin{subfigure}{0.495\textwidth}
		\includegraphics[width=\linewidth]{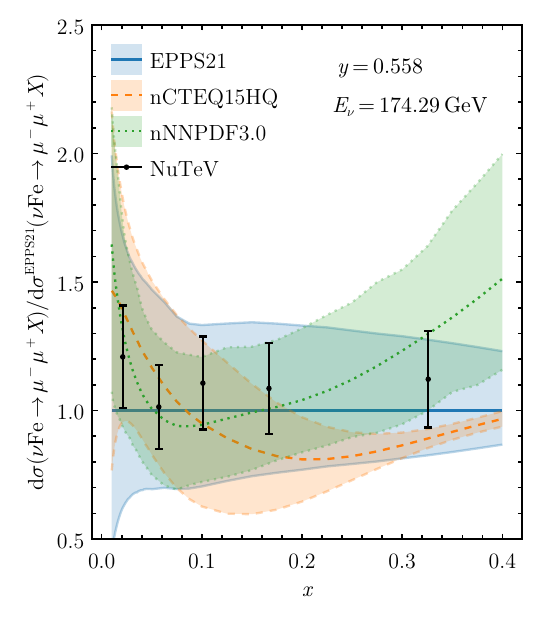}
	\end{subfigure}
	\begin{subfigure}{0.495\textwidth}
		\includegraphics[width=\linewidth]{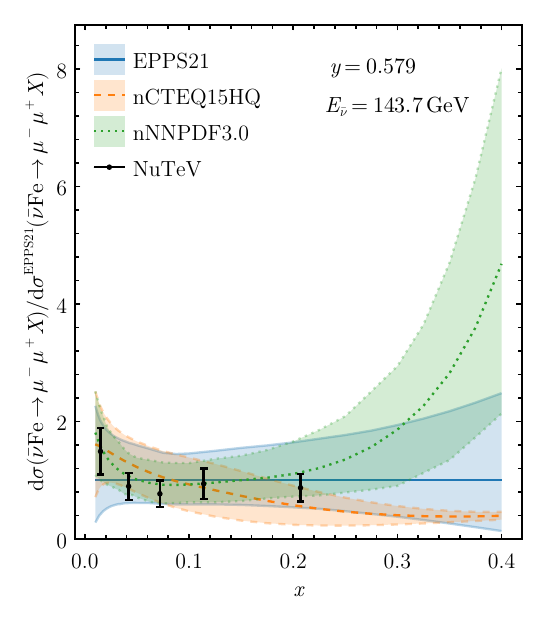}
	\end{subfigure}
	\caption{Same as in figure~\ref{fig:nutev_neutrino_pdf_comparison} (left) and figure~\ref{fig:nutev_antineutrino_pdf_comparison} (right), but normalized to the central \texttt{EPPS21} \cite{Eskola:2021nhw} value.}
	\label{fig:nutev_normalized}
\end{figure}

To study the differences between the three PDF sets in more detail, we show a representative example of the cross sections normalized to the central \texttt{EPPS21} value in figure~\ref{fig:nutev_normalized}. It particularly highlights the larger cross sections obtained with \texttt{nNNPDF3.0} at $0.3<x<0.4$, in contrast to \texttt{EPPS21} and \texttt{nCTEQ15HQ}. This behavior can again be understood from the strange-quark distributions in figure~\ref{fig:npdf_xs_10}. While these NuTeV data do not reach $x$ values much above $x=0.3$, the inclusive neutrino-nucleus data still extend to higher values of $x$ and particularly in the case of antineutrino scattering, the higher large-$x$ strange quarks of \texttt{nNNPDF3.0} are well visible also there, see e.g. figure~5 of ref.~\cite{Klasen:2023uqj}.

\begin{figure}[p!]
	\centering
	\includegraphics[width=\linewidth]{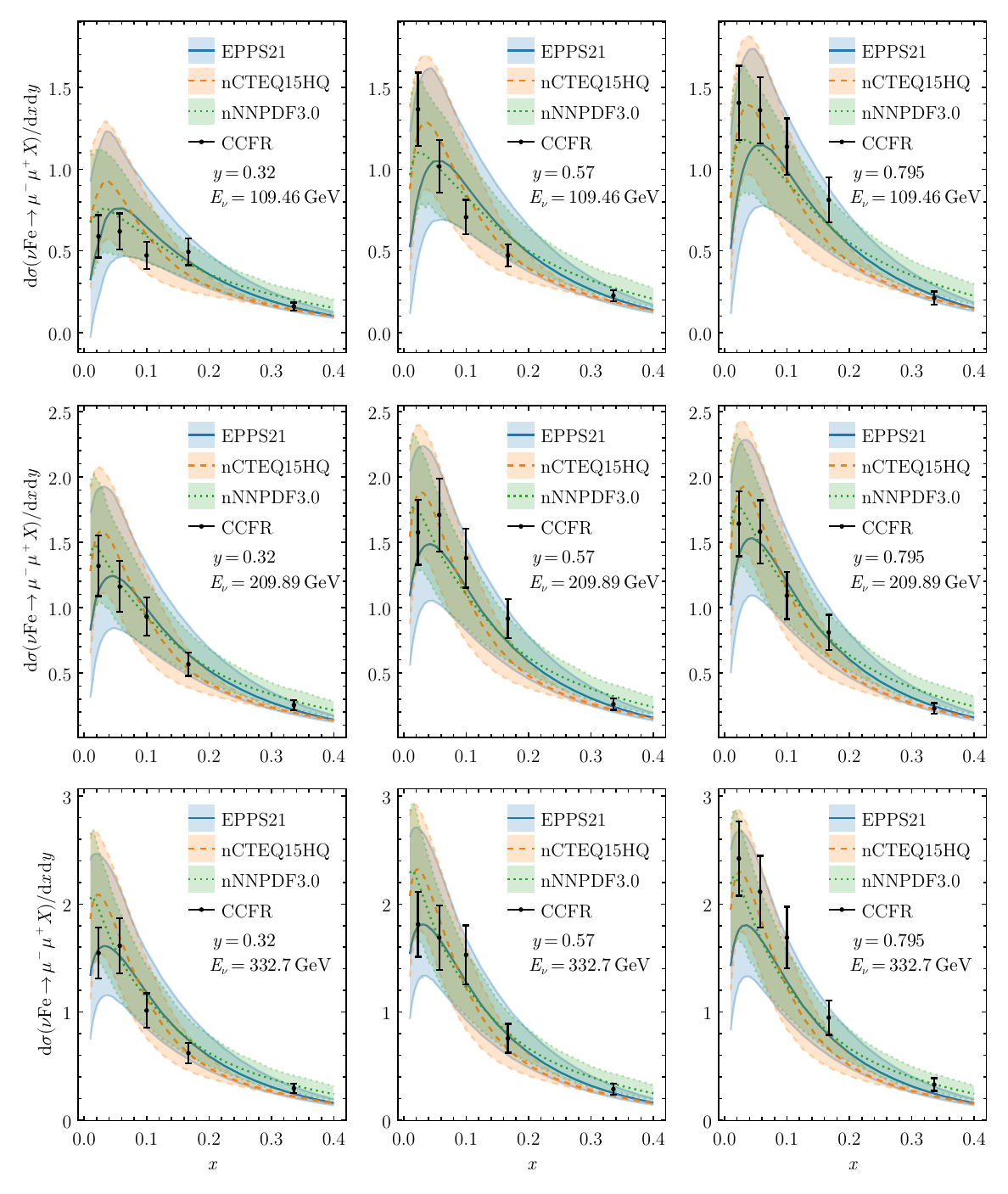}
	\caption{Same as figure~\ref{fig:nutev_neutrino_pdf_comparison}, but with CCFR kinematics and data from ref.~\cite{NuTeV:2001dfo}.}
	\label{fig:ccfr_neutrino_pdf_comparison}
\end{figure}

\begin{figure}[p!]
	\centering
	\includegraphics[width=\linewidth]{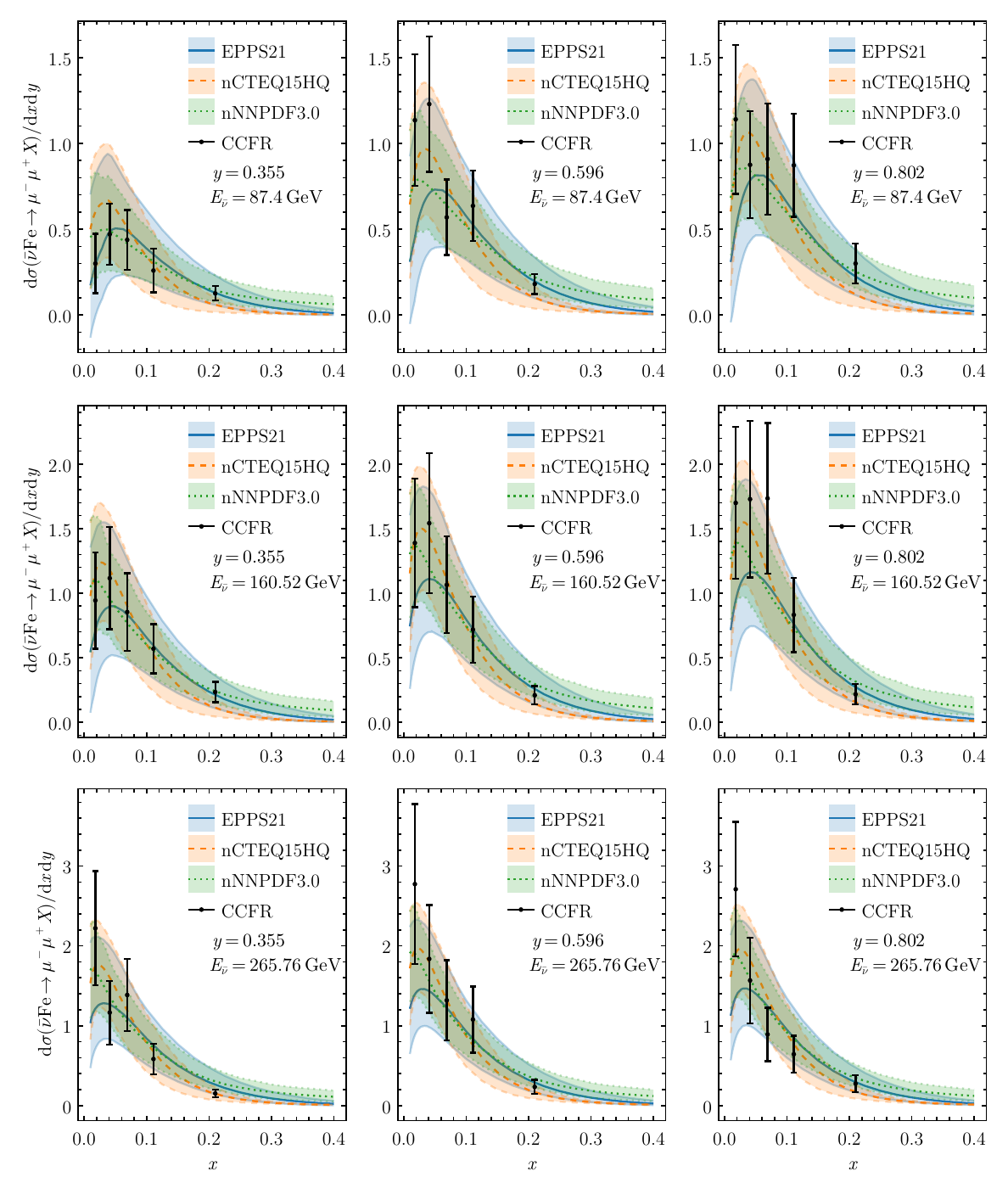}
	\caption{Same as figure~\ref{fig:ccfr_neutrino_pdf_comparison}, but with antineutrino scattering.}
	\label{fig:ccfr_antineutrino_pdf_comparison}
\end{figure}

Figures~\ref{fig:ccfr_neutrino_pdf_comparison} and \ref{fig:ccfr_antineutrino_pdf_comparison} show the analogous results for the older CCFR data for neutrino and antineutrino scattering, respectively. While the $y$ and $E_\nu$ bins are slightly different, the kinematic reach of these data is similar to the NuTeV data. In general, the agreement with the data and the calculations applying different nuclear PDF sets are similar as in the case of the NuTeV data, though the preferred PDF set at small-$x$ varies for different bins of $y$: the smallest $y$ data tend to be best described with \texttt{EPPS21}, but in the largest $y$ bins the larger cross sections provided by \texttt{nCTEQ15HQ} and \texttt{nNNPDF3.0} are preferred.

%%%%%%%%%%%%%%%%%%%%%%%%%%%%%%%%%%%%%%%%%%%%%%%%%%%%%%%%%%%%%%%%%%%%%%%%%%%%%%%%%%%%%%%%%%%%%%%%%%%%%%%%%%%

\subsection{Nuclear effects}

As the NuTeV and CCFR data have been taken with an iron target, nuclear effects are involved \cite{Klasen:2023uqj}. In fits of free-proton PDFs, one approach seen in the literature \cite{NuTeV:2005wsg,Hou:2019efy} (see also ref.~\cite{Accardi:2021ysh}) has been to model the nuclear correction as an overall multiplicative factor depending only on $x$, $f(x)$, instead of applying the correction on a flavor-by-flavor basis. For example, ref.~\cite{NuTeV:2005wsg} uses the so-called SLAC/NMC form,
\begin{equation}
\label{eq:nuclear_correction_parametrization}
f(x) = 1.10-0.36x-0.28e^{-21.94x}+2.77x^{14.42},
\end{equation}
which is found by fitting to charged-lepton neutral-current DIS processes \cite{Seligman:1997fe}. It should be noted that this correction was originally fitted to the ratio $F_2^A/F_2^D$ of the structure function $F_2$ from nuclear targets (iron and calcium) to deuterium.

We can calculate the isospin-corrected multiplicative nuclear effect $R$ by taking the ratio between the dimuon cross section computed with nuclear PDFs and the dimuon cross section computed with isospin-corrected proton PDFs
\begin{equation}
\label{eq:heavy_nuclear_correction}
	R\equiv \frac{\dd\sigma(\nu_\mu N\to \mu\mu X)|_{\text{nuclear PDF}}}{\dd\sigma(\nu_\mu N\to \mu\mu X)|_{\text{isospin PDF}}}.
\end{equation}
The isospin-corrected proton PDFs $f_A^p$ simply take the number of protons ($Z$) and neutrons ($N$) in a nucleus into account via isospin symmetry and are defined as
\begin{align}
	u_A^p(x, Q^2)&=\frac{Z}{A}u^p(x, Q^2)+\frac{N}{A}d^p(x, Q^2), \\
	d_A^p(x, Q^2)&=\frac{Z}{A}d^p(x, Q^2)+\frac{N}{A}u^p(x, Q^2),
\end{align}
where $f^p$ are the usual proton PDFs. The definition is analogous for $\overline u_A^p$ and $\overline d_A^p$. For other flavors and the gluon, $f_A^p(x, Q^2)=f^p(x, Q^2)$. 

\begin{figure}[h!]
	\begin{subfigure}{0.495\textwidth}
		\includegraphics[width=\linewidth]{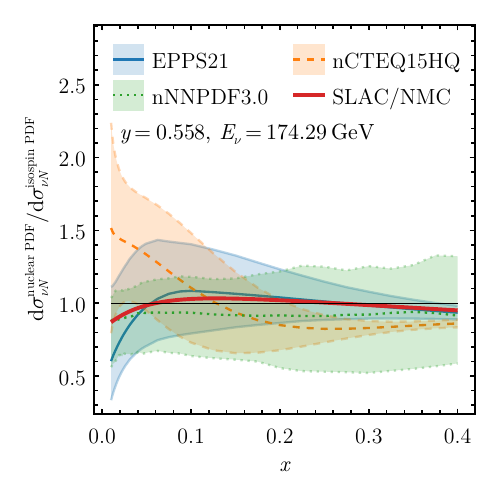}
	\end{subfigure}
	\begin{subfigure}{0.495\textwidth}
		\includegraphics[width=\linewidth]{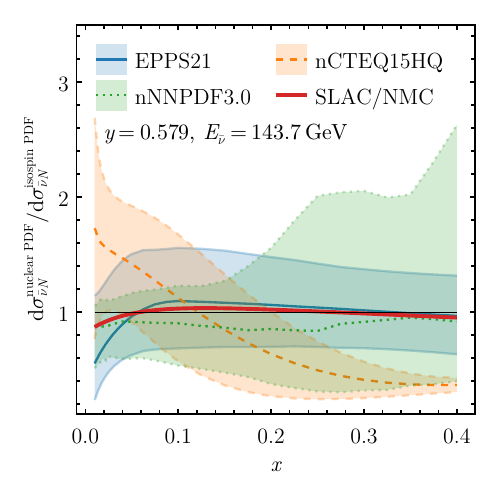}
	\end{subfigure}
	\caption{Isospin-corrected nuclear ratio, defined in eq.~\eqref{eq:heavy_nuclear_correction}, and evaluated using \texttt{EPPS21} \cite{Eskola:2021nhw}, \texttt{nCTEQ15HQ} \cite{Duwentaster:2022kpv}, and \texttt{nNNPDF3.0} \cite{AbdulKhalek:2022fyi} at NLO. The figure on the left shows the neutrino scattering case and the figure on the right shows the antineutrino case. The bands depict the PDF uncertainty at $\SI{90}{\percent}$ confidence level, which is calculated by correlating the uncertainties in the ratio. As a reference, the SLAC/NMC form \cite{Seligman:1997fe} of eq.~\eqref{eq:nuclear_correction_parametrization} is shown as the thicker red curve.}
	\label{fig:heavy_nucleus_correction}
\end{figure}

Figure~\ref{fig:heavy_nucleus_correction} shows the isospin-corrected nuclear ratio for neutrino and antineutrino scattering. For each nuclear PDF set, we use the corresponding baseline proton PDF set. For \texttt{EPPS21} \cite{Eskola:2021nhw}, this is \texttt{CT18ANLO} \cite{Hou:2019efy}, while \texttt{nCTEQ15HQ} \cite{Duwentaster:2022kpv} and \texttt{nNNPDF3.0} \cite{AbdulKhalek:2022fyi} come with their own baseline proton PDFs. The uncertainty is calculated by accounting for the correlations between the nuclear and ``isospin'' PDFs. While the three considered nuclear PDF sets mostly agree within their uncertainties,  there are some clear differences as well. The \texttt{EPPS21} result roughly follows eq.~\eqref{eq:nuclear_correction_parametrization}, where the correction is above unity in the region $0.05< x< 0.26$ (antishadowing) and below unity elsewhere (shadowing, EMC effect). On the other hand, the central curve of \texttt{nCTEQ15HQ} is above unity when $x<0.13$, but below it when $x>0.13$. Finally, \texttt{nNNPDF3.0} prefers to stay below unity in the entire $x$-range considered here. There are also some differences between neutrino and antineutrino scattering particularly in the case of \texttt{nCTEQ15HQ}: at $x=0.4$, the nuclear effect in the neutrino case is nearly double of that in the case of antineutrinos. The mutual behaviour of nuclear effects with different nuclear PDFs follows rather well the relative ordering of nuclear effects in the strange-quark distributions, see e.g. figure 4 of ref.~\cite{Klasen:2023uqj}. Overall, we conclude that the nuclear effects in the dimuon processes come with a significant uncertainty.

\subsection{Effective acceptance}

\begin{figure}[p!]
	\centering
	\includegraphics[width=\linewidth]{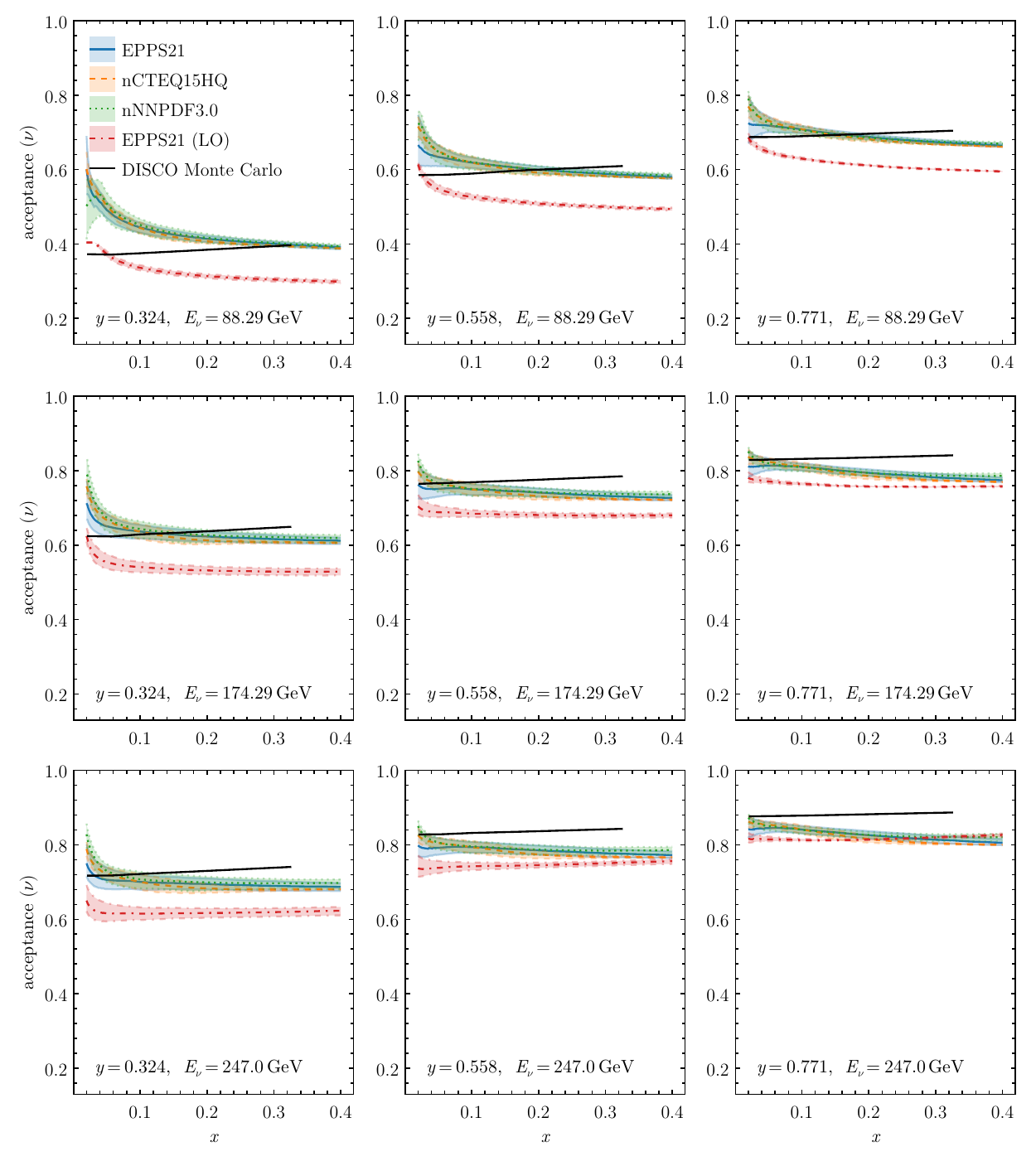}
	\caption{Acceptance correction, as defined in eq.~\eqref{eq:acceptance}, in neutrino scattering and in the NuTeV kinematics evaluated using the \texttt{EPPS21} \cite{Eskola:2021nhw}, \texttt{nCTEQ15HQ} \cite{Duwentaster:2022kpv}, and \texttt{nNNPDF3.0} \cite{AbdulKhalek:2022fyi} PDF sets. The calculation is done at NLO, except for the case explictly marked as LO. The uncertainty bands correspond to the PDF uncertainties at $\SI{90}{\percent}$ confidence level. Our calculation is compared against the DISCO Monte-Carlo calculation in ref.~\cite{Mason:2006qa}.}
	\label{fig:acceptance_nutev_neutrino}
\end{figure}

\begin{figure}[h!]
	\centering
	\includegraphics[width=\linewidth]{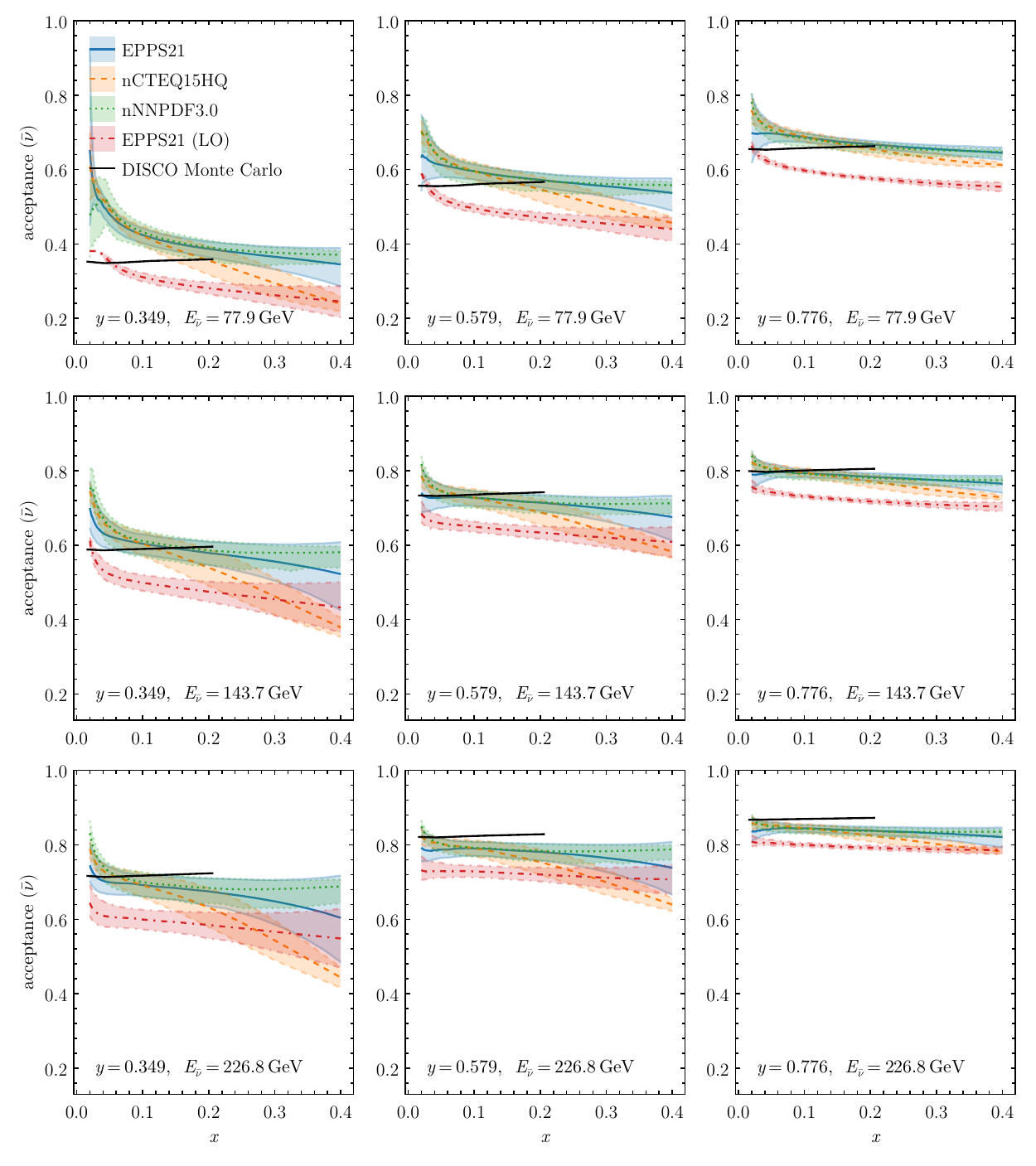}
	\caption{Same as in figure~\ref{fig:acceptance_nutev_neutrino}, but for antineutrino scattering.}
	\label{fig:acceptance_nutev_antineutrino}
\end{figure}

The framework we have adopted allows us to directly implement kinematical cuts for the final-state muons, but in order to compare our results to the factorized approach in eq.~\eqref{eq:charm_production_factorization}, we calculate the equivalent of the effective acceptance correction $\mathcal{A}$ from
\begin{equation}
\label{eq:acceptance}
	\mathcal{A}=\frac{1}{\mathcal{B}_\mu}\frac{\dd\sigma(\nu_\mu N\to \mu\mu X)}{\dd\sigma(\nu_\mu N\to \mu cX)},
\end{equation}
where $\mathcal{B}_\mu$ is the average semileptonic branching ratio of charm mesons. The denominator $\dd\sigma(\nu_\mu N\to \mu cX)$ corresponds to the inclusive charm production cross section and, as the numerator, it is calculated in the variable flavour-number scheme approximating the quark-mass effects with the slow-rescaling parameter in eq.~(\ref{eq:slowrescaling}). In addition to channels with a charm quark explicitly in the final state, it includes contributions in which the presence of the final-state charm in implicit, e.g. the partonic channel $\bar{c} \rightarrow \bar s$ at LO. The origin of such contributions are the diagrams in which an initial-state gluon splits into a $c\overline{c}$ pair producing DGLAP logarithms $\log(Q^2/m_c^2)$, which are resummed into the definition of the (anti)charm-quark PDFs. In the gluon-initiated semi-inclusive case, such logarithms do not appear \cite{Kretzer:1997pd}. Both cross sections in the ratio of eq.~(\ref{eq:acceptance}) are calculated consistently at the same perturbative order, i.e. either at LO or NLO.

It is not obvious what value one should use for $\mathcal{B}_\mu$. Some fits \cite{Hou:2019efy,NNPDF:2017mvq,Martin:2009iq} have used a value of $0.099 \pm 0.012$, which has been used by the CCFR collaboration \cite{CCFR:1994ikl}. The PDG value 
$0.087 \pm 0.005$ is an average of several measurements \cite{Workman:2022ynf}, with more values listed in ref.~\cite{Mason:2006qa}. It should be noted that the PDG value of $\left|V_{cd}\right|$ is also dependent on the value of $\mathcal{B}_\mu$. Several groups have moved away from the CCFR value for $\mathcal{B}_\mu$, see for example refs. \cite{Ball:2018twp,Harland-Lang:2014zoa} for more discussion. For figures~\ref{fig:acceptance_nutev_neutrino} and \ref{fig:acceptance_nutev_antineutrino}, we follow ref.~\cite{Harland-Lang:2014zoa} and use $0.0919\pm \SI{10}{\percent}$, which is based on FNAL E531 data \cite{Bolton:1997pq}. While the value of $\mathcal{B}_\mu$ can have a significant effect on the dimuon production cross-section values when using eq.~\eqref{eq:charm_production_factorization}, its impact in global fits can be mitigated by including the uncertainty of $\mathcal{B}_\mu$ as a correlated systematic uncertainty in the global analysis. 

In figures~\ref{fig:acceptance_nutev_neutrino} and \ref{fig:acceptance_nutev_antineutrino}, we compare the acceptance values, computed according to eq.~\eqref{eq:acceptance}, to those from a DISCO NLO Monte-Carlo calculation \cite{Mason:2006qa} for neutrino and antineutrino scattering, respectively. It should be noted that the $\SI{10}{\percent}$ normalization uncertainty of $\mathcal{B}_\mu$ in eq.~\eqref{eq:acceptance} is not included in figures~\ref{fig:acceptance_nutev_neutrino} and \ref{fig:acceptance_nutev_antineutrino}. While the NLO values of $\mathcal{A}$ from the DISCO Monte-Carlo calculation of ref.~\cite{Mason:2006qa} are roughly in line with the values computed here directly, there are clear systematic differences which are roughly at the $10\,\%$ level. Whereas our calculation shows a general trend of decreasing acceptance with increasing $x$, the DISCO Monte-Carlo calculation tends to have a modest rise with increasing $x$. Furthermore, we notice that for small values of $y$ and $E$, our results tend to be above the DISCO calculation, but with increasing $y$ and $E$ the ordering goes the other way around. Thus it is clear that a single value of $\mathcal{B}_\mu$ is not enough to match our calculation to the DISCO one. The acceptances calculated with different nuclear PDFs are in a rather good mutual agreement with all sharing the same systematic behaviour, but especially in the antineutrino case shown in figure~\ref{fig:acceptance_nutev_antineutrino} there are larger uncertainties originating from the nuclear PDFs. In addition to the NLO results, figures~\ref{fig:acceptance_nutev_neutrino} and \ref{fig:acceptance_nutev_antineutrino} also show the LO acceptance correction, where both the dimuon production and inclusive charm production cross sections are calculated at LO. As can be seen, the LO and NLO acceptances differ quite significantly. Based on these comparisons, we see that the acceptances in eq.~\eqref{eq:charm_production_factorization} depend both on the nuclear PDFs as well as on the perturbative order of the calculation.

%%%%%%%%%%%%%%%%%%%%%%%%%%%%%%%%%%%%%%%%%%%%%%%%%%%%%%%%%%%%%%%%%%%%%%%%%%%%%%%
\section{Conclusion and outlook}
\label{sec:Conclusion}

We have presented an NLO calculation of dimuon production in semi-inclusive deep-inelastic neutrino-nucleus scattering. In comparison to how this process is usually calculated in PDF fits, our calculation involves scale-dependent FFs to turn final-state charm quarks into charmed hadrons while at the same time resumming collinear final-state radiation, and a decay function fitted to $e^+e^-$ data to produce muons from the charmed-hadron decays. The major advantage of our approach is that we do not need to rely on the approximative, factorized acceptance and branching-ratio corrections to account for the experimental cuts on the final-state muon energies. 
Instead, the kinematical cuts are a natural ingredient of the calculation.
The perturbative accuracy of the presented framework can be systematically improved by including higher-order corrections to the coefficient functions and employing PDFs and FFs evolving with higher-order splitting functions. The approach is also generalizable to other processes where an intermediate-state particle decays into a given particle.

We assessed different theoretical uncertainties arising from the decay function, parton distribution functions, and variations in the renormalization, factorization, and fragmentation scales. While the uncertainties in the decay-function fit translated into few-percent differences in the dimuon cross sections, the NLO scale uncertainty was of the order of $10\,\%$ in the perturbative regime. We also studied the size of the the approximative NNLO corrections in the coefficient functions and found them to be comparable to the size of the NLO scale uncertainty. By far, the largest uncertainty was associated with the strange-quark component of nuclear PDFs, which could be $\sim 100\,\%$ in the data-constrained region. This indicates that there is a potential for broader use of these data in global fits of nuclear PDFs. 

We also studied the effective acceptance and nuclear effects which are often taken as inputs in free-proton fits to these dimuon data. As for the effective acceptance we found systematic differences of the order of $10\,\%$ between our calculation and widely applied acceptance corrections derived from the DISCO Monte-Carlo framework. Moreover, the effective acceptance is clearly dependent on the perturbative order and --- particularly in the case of antineutrino scattering --- also PDF dependent. This indicates that using an external acceptance correction can, perhaps, lead to some biases in PDF fits. As for the nuclear correction, the latest globally analyzed nuclear PDFs led to quite significant mutual differences with qualitatively distinct behaviour and significant uncertainties. 

In the near future, we plan to improve the framework introduced here in several ways. First, it can be straightforwardly extended to full GM-VFNS, thereby accounting also for the subleading effects of the charm-quark mass. Second, the calculations for the full massless NNLO SIDIS coefficient functions in the photon exchange have been very recently completed and thus we expect the full charged-current coefficient functions to become available soon. Third, similarly to the case of inclusive neutrino DIS, electroweak radiative and target-mass corrections can be included in the SIDIS cross sections as well. Eventually, it will be interesting to see whether all this will bring insights to the observed tensions between dimuon production in neutrino DIS and heavy gauge boson production at the LHC.  

\section*{Acknowledgements}

We acknowledge the financial support from the Magnus Ehrnrooth foundation (S.Y.), the Research Council of Finland Project No. 331545 (I.H.), and the Center of Excellence in Quark Matter of the Research Council of Finland, project 346326. The reported work is associated with the European Research Council project ERC-2018-ADG-835105 YoctoLHC. We acknowledge grants of computer capacity from the Finnish Grid and Cloud Infrastructure (persistent identifier
   \texttt{urn:nbn:fi:research-infras-2016072533}) and the Finnish IT Center for Science (CSC), under the project jyy2580. 

\bibliographystyle{JHEP}
\bibliography{refs.bib}

\end{document}